

\catcode`\@=11
\font\tensmc=cmcsc10      
\def\smc{\tensmc}

\def\hcorrection#1{\advance\hoffset by #1 }
\def\vcorrection#1{\advance\voffset by #1 }
\def\wlog#1{}
\newif\iftitle@
\outer\def\title{\title@true\vglue 24\p@ plus 12\p@ minus 12\p@
   \bgroup\let\\=\cr\tabskip\centering
   \halign to \hsize\bgroup\tenbf\hfill\ignorespaces##\unskip\hfill\cr}
\def\endtitle{\cr\egroup\egroup\vglue 18\p@ plus 12\p@ minus 6\p@}
\outer\def\author{\iftitle@\vglue -18\p@ plus -12\p@ minus -6\p@\fi\vglue
    12\p@ plus 6\p@ minus 3\p@\bgroup\let\\=\cr\tabskip\centering
    \halign to \hsize\bgroup\smc\hfill\ignorespaces##\unskip\hfill\cr}
\def\endauthor{\cr\egroup\egroup\vglue 18\p@ plus 12\p@ minus 6\p@}
\outer\def\heading{\bigbreak\bgroup\let\\=\cr\tabskip\centering
    \halign to \hsize\bgroup\smc\hfill\ignorespaces##\unskip\hfill\cr}
\def\endheading{\cr\egroup\egroup\nobreak\medskip}

\outer\def\endproclaim{\par\ifdim\lastskip<\medskipamount\removelastskip
  \penalty 55 \fi\medskip\rm}
\outer\def\demo#1{\par\ifdim\lastskip<\smallskipamount\removelastskip
    \smallskip\fi\noindent{\smc\ignorespaces#1\unskip:\enspace}\rm
      \ignorespaces}

\newcount\footmarkcount@
\footmarkcount@=1
\def\makefootnote@#1#2{\insert\footins{\interlinepenalty=100
  \splittopskip=\ht\strutbox \splitmaxdepth=\dp\strutbox
  \floatingpenalty=\@MM
  \leftskip=\z@\rightskip=\z@\spaceskip=\z@\xspaceskip=\z@
  \noindent{#1}\footstrut\rm\ignorespaces #2\strut}}
\def\footnote{\let\@sf=\empty\ifhmode\edef\@sf{\spacefactor
   =\the\spacefactor}\/\fi\futurelet\next\footnote@}
\def\footnote@{\ifx"\next\let\next\footnote@@\else
    \let\next\footnote@@@\fi\next}
\def\footnote@@"#1"#2{#1\@sf\relax\makefootnote@{#1}{#2}}
\def\footnote@@@#1{$^{\number\footmarkcount@}$\makefootnote@
   {$^{\number\footmarkcount@}$}{#1}\global\advance\footmarkcount@ by 1 }

\hyphenation{man-u-script man-u-scripts ap-pen-dix ap-pen-di-ces}
\hyphenation{data-base data-bases}
\ifx\amstexloaded@\relax\catcode`\@=13
  \endinput\else\let\amstexloaded@=\relax\fi
\newlinechar=`\^^J
\def\eat@#1{}
\def\Space@.{\futurelet\Space@\relax}
\Space@. %
\newhelp\athelp@
{Only certain combinations beginning with @ make sense to me.^^J
Perhaps you wanted \string\@\space for a printed @?^^J
I've ignored the character or group after @.}
\def\futureletnextat@{\futurelet\next\at@}
{\catcode`\@=\active
\lccode`\Z=`\@ \lowercase
{\gdef@{\expandafter\csname futureletnextatZ\endcsname}
\expandafter\gdef\csname atZ\endcsname
   {\ifcat\noexpand\next a\def\next{\csname atZZ\endcsname}\else
   \ifcat\noexpand\next0\def\next{\csname atZZ\endcsname}\else
    \def\next{\csname atZZZ\endcsname}\fi\fi\next}
\expandafter\gdef\csname atZZ\endcsname#1{\expandafter
   \ifx\csname #1Zat\endcsname\relax\def\next
     {\errhelp\expandafter=\csname athelpZ\endcsname
      \errmessage{Invalid use of \string@}}\else
       \def\next{\csname #1Zat\endcsname}\fi\next}
\expandafter\gdef\csname atZZZ\endcsname#1{\errhelp
    \expandafter=\csname athelpZ\endcsname
      \errmessage{Invalid use of \string@}}}}
\def\atdef@#1{\expandafter\def\csname #1@at\endcsname}
\newhelp\defahelp@{If you typed \string\define\space cs instead of
\string\define\string\cs\space^^J
I've substituted an inaccessible control sequence so that your^^J
definition will be completed without mixing me up too badly.^^J
If you typed \string\define{\string\cs} the inaccessible control sequence^^J
was defined to be \string\cs, and the rest of your^^J
definition appears as input.}
\newhelp\defbhelp@{I've ignored your definition, because it might^^J
conflict with other uses that are important to me.}
\def\define{\futurelet\next\define@}
\def\define@{\ifcat\noexpand\next\relax
  \def\next{\define@@}%
  \else\errhelp=\defahelp@
  \errmessage{\string\define\space must be followed by a control
     sequence}\def\next{\def\garbage@}\fi\next}
\def\undefined@{}
\def\preloaded@{}
\def\define@@#1{\ifx#1\relax\errhelp=\defbhelp@
   \errmessage{\string#1\space is already defined}\def\next{\def\garbage@}%
   \else\expandafter\ifx\csname\expandafter\eat@\string
         #1@\endcsname\undefined@\errhelp=\defbhelp@
   \errmessage{\string#1\space can't be defined}\def\next{\def\garbage@}%
   \else\expandafter\ifx\csname\expandafter\eat@\string#1\endcsname\relax
     \def\next{\def#1}\else\errhelp=\defbhelp@
     \errmessage{\string#1\space is already defined}\def\next{\def\garbage@}%
      \fi\fi\fi\next}
\def\famzero{\fam\z@}

\def\exp{\mathop{\famzero exp}\nolimits}

\def\lim{\mathop{\famzero lim}}

\def\sinh{\mathop{\famzero sinh}\nolimits}

\def\textfont@#1#2{\def#1{\relax\ifmmode
    \errmessage{Use \string#1\space only in text}\else#2\fi}}
\textfont@\rm\tenrm
\textfont@\it\tenit
\textfont@\sl\tensl
\textfont@\bf\tenbf
\textfont@\smc\tensmc
\let\ic@=\/
\def\/{\unskip\ic@}
\def\textfonti{\the\textfont1 }
\def\t#1#2{{\edef\next{\the\font}\textfonti\accent"7F \next#1#2}}
\let\B=\=
\let\D=\.
\def~{\unskip\nobreak\ \ignorespaces}
{\catcode`\@=\active
\gdef\@{\char'100 }}
\atdef@-{\leavevmode\futurelet\next\athyph@}
\def\athyph@{\ifx\next-\let\next=\athyph@@
  \else\let\next=\athyph@@@\fi\next}
\def\athyph@@@{\hbox{-}}
\def\athyph@@#1{\futurelet\next\athyph@@@@}
\def\athyph@@@@{\if\next-\def\next##1{\hbox{---}}\else
    \def\next{\hbox{--}}\fi\next}
\def\.{.\spacefactor=\@m}
\atdef@.{\null.}
\atdef@,{\null,}
\atdef@;{\null;}
\atdef@:{\null:}
\atdef@?{\null?}
\atdef@!{\null!}
\def\srdr@{\thinspace}
\def\drsr@{\kern.02778em}
\def\sldl@{\kern.02778em}
\def\dlsl@{\thinspace}
\atdef@"{\unskip\futurelet\next\atqq@}
\def\atqq@{\ifx\next\Space@\def\next. {\atqq@@}\else
         \def\next.{\atqq@@}\fi\next.}
\def\atqq@@{\futurelet\next\atqq@@@}
\def\atqq@@@{\ifx\next`\def\next`{\atqql@}\else\def\next'{\atqqr@}\fi\next}
\def\atqql@{\futurelet\next\atqql@@}
\def\atqql@@{\ifx\next`\def\next`{\sldl@``}\else\def\next{\dlsl@`}\fi\next}
\def\atqqr@{\futurelet\next\atqqr@@}
\def\atqqr@@{\ifx\next'\def\next'{\srdr@''}\else\def\next{\drsr@'}\fi\next}

\def\textfontii{\the\textfont2 }
\def\{{\relax\ifmmode\lbrace\else
    {\textfontii f}\spacefactor=\@m\fi}
\def\}{\relax\ifmmode\rbrace\else
    \let\@sf=\empty\ifhmode\edef\@sf{\spacefactor=\the\spacefactor}\fi
      {\textfontii g}\@sf\relax\fi}
\def\nonhmodeerr@#1{\errmessage
     {\string#1\space allowed only within text}}
\def\linebreak{\relax\ifhmode\unskip\break\else
    \nonhmodeerr@\linebreak\fi}
\def\allowlinebreak{\relax
   \ifhmode\allowbreak\else\nonhmodeerr@\allowlinebreak\fi}
\newskip\saveskip@
\def\nolinebreak{\relax\ifhmode\saveskip@=\lastskip\unskip
  \nobreak\ifdim\saveskip@>\z@\hskip\saveskip@\fi
   \else\nonhmodeerr@\nolinebreak\fi}
\def\newline{\relax\ifhmode\null\hfil\break
    \else\nonhmodeerr@\newline\fi}
\def\nonmathaerr@#1{\errmessage
     {\string#1\space is not allowed in display math mode}}
\def\nonmathberr@#1{\errmessage{\string#1\space is allowed only in math mode}}
\def\mathbreak{\relax\ifmmode\ifinner\break\else
   \nonmathaerr@\mathbreak\fi\else\nonmathberr@\mathbreak\fi}
\def\nomathbreak{\relax\ifmmode\ifinner\nobreak\else
    \nonmathaerr@\nomathbreak\fi\else\nonmathberr@\nomathbreak\fi}
\def\allowmathbreak{\relax\ifmmode\ifinner\allowbreak\else
     \nonmathaerr@\allowmathbreak\fi\else\nonmathberr@\allowmathbreak\fi}
\def\pagebreak{\relax\ifmmode
   \ifinner\errmessage{\string\pagebreak\space
     not allowed in non-display math mode}\else\postdisplaypenalty-\@M\fi
   \else\ifvmode\penalty-\@M\else\edef\spacefactor@
       {\spacefactor=\the\spacefactor}\vadjust{\penalty-\@M}\spacefactor@
        \relax\fi\fi}
\def\nopagebreak{\relax\ifmmode
     \ifinner\errmessage{\string\nopagebreak\space
    not allowed in non-display math mode}\else\postdisplaypenalty\@M\fi
    \else\ifvmode\nobreak\else\edef\spacefactor@
        {\spacefactor=\the\spacefactor}\vadjust{\penalty\@M}\spacefactor@
         \relax\fi\fi}
\def\newpage{\relax\ifvmode\vfill\penalty-\@M\else\nonvmodeerr@\newpage\fi}
\def\nonvmodeerr@#1{\errmessage
    {\string#1\space is allowed only between paragraphs}}
\def\smallpagebreak{\relax\ifvmode\smallbreak
      \else\nonvmodeerr@\smallpagebreak\fi}
\def\medpagebreak{\relax\ifvmode\medbreak
       \else\nonvmodeerr@\medpagebreak\fi}
\def\bigpagebreak{\relax\ifvmode\bigbreak
      \else\nonvmodeerr@\bigpagebreak\fi}
\newdimen\captionwidth@
\captionwidth@=\hsize
\advance\captionwidth@ by -1.5in
\def\caption#1{}
\def\topspace#1{\gdef\thespace@{#1}\ifvmode\def\next
    {\futurelet\next\topspace@}\else\def\next{\nonvmodeerr@\topspace}\fi\next}
\def\topspace@{\ifx\next\Space@\def\next. {\futurelet\next\topspace@@}\else
     \def\next.{\futurelet\next\topspace@@}\fi\next.}
\def\topspace@@{\ifx\next\caption\let\next\topspace@@@\else
    \let\next\topspace@@@@\fi\next}
 \def\topspace@@@@{\topinsert\vbox to
       \thespace@{}\endinsert}
\def\topspace@@@\caption#1{\topinsert\vbox to
    \thespace@{}\nobreak
      \smallskip
    \setbox\z@=\hbox{\noindent\ignorespaces#1\unskip}%
   \ifdim\wd\z@>\captionwidth@
   \centerline{\vbox{\hsize=\captionwidth@\noindent\ignorespaces#1\unskip}}%
   \else\centerline{\box\z@}\fi\endinsert}
\def\midspace#1{\gdef\thespace@{#1}\ifvmode\def\next
    {\futurelet\next\midspace@}\else\def\next{\nonvmodeerr@\midspace}\fi\next}
\def\midspace@{\ifx\next\Space@\def\next. {\futurelet\next\midspace@@}\else
     \def\next.{\futurelet\next\midspace@@}\fi\next.}
\def\midspace@@{\ifx\next\caption\let\next\midspace@@@\else
    \let\next\midspace@@@@\fi\next}
 \def\midspace@@@@{\midinsert\vbox to
       \thespace@{}\endinsert}
\def\midspace@@@\caption#1{\midinsert\vbox to
    \thespace@{}\nobreak
      \smallskip
      \setbox\z@=\hbox{\noindent\ignorespaces#1\unskip}%
      \ifdim\wd\z@>\captionwidth@
    \centerline{\vbox{\hsize=\captionwidth@\noindent\ignorespaces#1\unskip}}%
    \else\centerline{\box\z@}\fi\endinsert}
\mathchardef\prime@="0230
\def\prime{{{}\prime@{}}}
\def\prim@s{\prime@\futurelet\next\pr@m@s}

\def\,{\relax\ifmmode\mskip\thinmuskip\else\thinspace\fi}
\def\!{\relax\ifmmode\mskip-\thinmuskip\else\negthinspace\fi}
\def\frac#1#2{{#1\over#2}}

\def\:{\nobreak\hskip.1111em{:}\hskip.3333em plus .0555em\relax}
\def\intic@{\mathchoice{\hskip5\p@}{\hskip4\p@}{\hskip4\p@}{\hskip4\p@}}
\def\negintic@
 {\mathchoice{\hskip-5\p@}{\hskip-4\p@}{\hskip-4\p@}{\hskip-4\p@}}
\def\intkern@{\mathchoice{\!\!\!}{\!\!}{\!\!}{\!\!}}
\def\intdots@{\mathchoice{\cdots}{{\cdotp}\mkern1.5mu
    {\cdotp}\mkern1.5mu{\cdotp}}{{\cdotp}\mkern1mu{\cdotp}\mkern1mu
      {\cdotp}}{{\cdotp}\mkern1mu{\cdotp}\mkern1mu{\cdotp}}}
\newcount\intno@
\def\iint{\intno@=\tw@\futurelet\next\ints@}
\def\iiint{\intno@=\thr@@\futurelet\next\ints@}
\def\iiiint{\intno@=4 \futurelet\next\ints@}
\def\idotsint{\intno@=\z@\futurelet\next\ints@}
\def\ints@{\findlimits@\ints@@}
\newif\iflimtoken@
\newif\iflimits@
\def\findlimits@{\limtoken@false\limits@false\ifx\next\limits
 \limtoken@true\limits@true\else\ifx\next\nolimits\limtoken@true\limits@false
    \fi\fi}
\def\multintlimits@{\intop\ifnum\intno@=\z@\intdots@
  \else\intkern@\fi
    \ifnum\intno@>\tw@\intop\intkern@\fi
     \ifnum\intno@>\thr@@\intop\intkern@\fi\intop}
\def\multint@{\int\ifnum\intno@=\z@\intdots@\else\intkern@\fi
   \ifnum\intno@>\tw@\int\intkern@\fi
    \ifnum\intno@>\thr@@\int\intkern@\fi\int}
\def\ints@@{\iflimtoken@\def\ints@@@{\iflimits@
   \negintic@\mathop{\intic@\multintlimits@}\limits\else
    \multint@\nolimits\fi\eat@}\else
     \def\ints@@@{\multint@\nolimits}\fi\ints@@@}
\def\Sb{_\bgroup\vspace@
        \baselineskip=\fontdimen10 \scriptfont\tw@
        \advance\baselineskip by \fontdimen12 \scriptfont\tw@
        \lineskip=\thr@@\fontdimen8 \scriptfont\thr@@
        \lineskiplimit=\thr@@\fontdimen8 \scriptfont\thr@@
        \Let@\vbox\bgroup\halign\bgroup \hfil$\scriptstyle
            {##}$\hfil\cr}
\def\endSb{\crcr\egroup\egroup\egroup}
\def\Sp{^\bgroup\vspace@
        \baselineskip=\fontdimen10 \scriptfont\tw@
        \advance\baselineskip by \fontdimen12 \scriptfont\tw@
        \lineskip=\thr@@\fontdimen8 \scriptfont\thr@@
        \lineskiplimit=\thr@@\fontdimen8 \scriptfont\thr@@
        \Let@\vbox\bgroup\halign\bgroup \hfil$\scriptstyle
            {##}$\hfil\cr}
\def\endSp{\crcr\egroup\egroup\egroup}
\def\Let@{\relax\iffalse{\fi\let\\=\cr\iffalse}\fi}
\def\vspace@{\def\vspace##1{\noalign{\vskip##1 }}}
\def\aligned{\,\vcenter\bgroup\vspace@\Let@\openup\jot\m@th\ialign
  \bgroup \strut\hfil$\displaystyle{##}$&$\displaystyle{{}##}$\hfil\crcr}
\def\endaligned{\crcr\egroup\egroup}
\def\matrix{\,\vcenter\bgroup\Let@\vspace@
    \normalbaselines
  \m@th\ialign\bgroup\hfil$##$\hfil&&\quad\hfil$##$\hfil\crcr
    \mathstrut\crcr\noalign{\kern-\baselineskip}}
\def\endmatrix{\crcr\mathstrut\crcr\noalign{\kern-\baselineskip}\egroup
                \egroup\,}
\newtoks\hashtoks@
\hashtoks@={#}
\def\format{\crcr\egroup\iffalse{\fi\ifnum`}=0 \fi\format@}
\def\format@#1\\{\def\preamble@{#1}%
  \def\c{\hfil$\the\hashtoks@$\hfil}%
  \def\r{\hfil$\the\hashtoks@$}%
  \def\l{$\the\hashtoks@$\hfil}%
  \setbox\z@=\hbox{\xdef\Preamble@{\preamble@}}\ifnum`{=0 \fi\iffalse}\fi
   \ialign\bgroup\span\Preamble@\crcr}

\def\cases{\left\{\,\vcenter\bgroup\vspace@
     \normalbaselines\openup\jot\m@th
       \Let@\ialign\bgroup$##$\hfil&\quad$##$\hfil\crcr
      \mathstrut\crcr\noalign{\kern-\baselineskip}}

\newif\iftagsleft@
\tagsleft@true
\def\TagsOnRight{\global\tagsleft@false}
\def\tag#1$${\iftagsleft@\leqno\else\eqno\fi
 \hbox{\def\pagebreak{\global\postdisplaypenalty-\@M}%
 \def\nopagebreak{\global\postdisplaypenalty\@M}\rm(#1\unskip)}%
  $$\postdisplaypenalty\z@\ignorespaces}
\interdisplaylinepenalty=\@M
\def\allowdisplaybreak@{\def\allowdisplaybreak{\noalign{\allowbreak}}}
\def\displaybreak@{\def\displaybreak{\noalign{\break}}}
\def\align#1\endalign{\def\tag{&}\vspace@\allowdisplaybreak@\displaybreak@
  \iftagsleft@\lalign@#1\endalign\else
   \ralign@#1\endalign\fi}
\def\ralign@#1\endalign{\displ@y\Let@\tabskip\centering\halign to\displaywidth
     {\hfil$\displaystyle{##}$\tabskip=\z@&$\displaystyle{{}##}$\hfil
       \tabskip=\centering&\llap{\hbox{(\rm##\unskip)}}\tabskip\z@\crcr
             #1\crcr}}
\def\lalign@
 #1\endalign{\displ@y\Let@\tabskip\centering\halign to \displaywidth
   {\hfil$\displaystyle{##}$\tabskip=\z@&$\displaystyle{{}##}$\hfil
   \tabskip=\centering&\kern-\displaywidth
        \rlap{\hbox{(\rm##\unskip)}}\tabskip=\displaywidth\crcr
               #1\crcr}}
\def\overrightarrow{\mathpalette\overrightarrow@}
\def\overrightarrow@#1#2{\vbox{\ialign{$##$\cr
    #1{-}\mkern-6mu\cleaders\hbox{$#1\mkern-2mu{-}\mkern-2mu$}\hfill
     \mkern-6mu{\to}\cr
     \noalign{\kern -1\p@\nointerlineskip}
     \hfil#1#2\hfil\cr}}}
\def\overleftarrow{\mathpalette\overleftarrow@}
\def\overleftarrow@#1#2{\vbox{\ialign{$##$\cr
     #1{\leftarrow}\mkern-6mu\cleaders\hbox{$#1\mkern-2mu{-}\mkern-2mu$}\hfill
      \mkern-6mu{-}\cr
     \noalign{\kern -1\p@\nointerlineskip}
     \hfil#1#2\hfil\cr}}}
\def\overleftrightarrow{\mathpalette\overleftrightarrow@}
\def\overleftrightarrow@#1#2{\vbox{\ialign{$##$\cr
     #1{\leftarrow}\mkern-6mu\cleaders\hbox{$#1\mkern-2mu{-}\mkern-2mu$}\hfill
       \mkern-6mu{\to}\cr
    \noalign{\kern -1\p@\nointerlineskip}
      \hfil#1#2\hfil\cr}}}
\def\underrightarrow{\mathpalette\underrightarrow@}
\def\underrightarrow@#1#2{\vtop{\ialign{$##$\cr
    \hfil#1#2\hfil\cr
     \noalign{\kern -1\p@\nointerlineskip}
    #1{-}\mkern-6mu\cleaders\hbox{$#1\mkern-2mu{-}\mkern-2mu$}\hfill
     \mkern-6mu{\to}\cr}}}
\def\underleftarrow{\mathpalette\underleftarrow@}
\def\underleftarrow@#1#2{\vtop{\ialign{$##$\cr
     \hfil#1#2\hfil\cr
     \noalign{\kern -1\p@\nointerlineskip}
     #1{\leftarrow}\mkern-6mu\cleaders\hbox{$#1\mkern-2mu{-}\mkern-2mu$}\hfill
      \mkern-6mu{-}\cr}}}
\def\underleftrightarrow{\mathpalette\underleftrightarrow@}
\def\underleftrightarrow@#1#2{\vtop{\ialign{$##$\cr
      \hfil#1#2\hfil\cr
    \noalign{\kern -1\p@\nointerlineskip}
     #1{\leftarrow}\mkern-6mu\cleaders\hbox{$#1\mkern-2mu{-}\mkern-2mu$}\hfill
       \mkern-6mu{\to}\cr}}}
\def\sqrt#1{\radical"270370 {#1}}
\def\dots{\relax\ifmmode\let\next=\ldots\else\let\next=\tdots@\fi\next}
\def\tdots@{\unskip\ \tdots@@}
\def\tdots@@{\futurelet\next\tdots@@@}
\def\tdots@@@{$\mathinner{\ldotp\ldotp\ldotp}\,
   \ifx\next,$\else
   \ifx\next.\,$\else
   \ifx\next;\,$\else
   \ifx\next:\,$\else
   \ifx\next?\,$\else
   \ifx\next!\,$\else
   $ \fi\fi\fi\fi\fi\fi}
\def\text{\relax\ifmmode\let\next=\text@\else\let\next=\text@@\fi\next}
\def\text@@#1{\hbox{#1}}
\def\text@#1{\mathchoice
 {\hbox{\everymath{\displaystyle}\def\textfonti{\the\textfont1 }%
    \def\textfontii{\the\textfont2 }\textdef@@ T#1}}
 {\hbox{\everymath{\textstyle}\def\textfonti{\the\textfont1 }%
    \def\textfontii{\the\textfont2 }\textdef@@ T#1}}
 {\hbox{\everymath{\scriptstyle}\def\textfonti{\the\scriptfont1 }%
   \def\textfontii{\the\scriptfont2 }\textdef@@ S\rm#1}}
 {\hbox{\everymath{\scriptscriptstyle}\def\textfonti{\the\scriptscriptfont1 }%
   \def\textfontii{\the\scriptscriptfont2 }\textdef@@ s\rm#1}}}
\def\textdef@@#1{\textdef@#1\rm \textdef@#1\bf
   \textdef@#1\sl \textdef@#1\it}

\def\textdef@#1#2{\def\next{\csname\expandafter\eat@\string#2fam\endcsname}%
\if S#1\edef#2{\the\scriptfont\next\relax}%
 \else\if s#1\edef#2{\the\scriptscriptfont\next\relax}%
 \else\edef#2{\the\textfont\next\relax}\fi\fi}
\scriptfont\itfam=\tenit \scriptscriptfont\itfam=\tenit
\scriptfont\slfam=\tensl \scriptscriptfont\slfam=\tensl
\mathcode`\0="0030
\mathcode`\1="0031
\mathcode`\2="0032
\mathcode`\3="0033
\mathcode`\4="0034
\mathcode`\5="0035
\mathcode`\6="0036
\mathcode`\7="0037
\mathcode`\8="0038
\mathcode`\9="0039
\def\Cal{\relax\ifmmode\let\next=\Cal@\else
     \def\next{\errmessage{Use \string\Cal\space only in math mode}}\fi\next}
\def\Cal@#1{{\fam2 #1}}
\def\bold{\relax\ifmmode\let\next=\bold@\else
   \def\next{\errmessage{Use \string\bold\space only in math
      mode}}\fi\next}\def\bold@#1{{\fam\bffam #1}}
\mathchardef\Gamma="0000
\mathchardef\Delta="0001
\mathchardef\Theta="0002
\mathchardef\Lambda="0003
\mathchardef\Xi="0004
\mathchardef\Pi="0005
\mathchardef\Sigma="0006
\mathchardef\Upsilon="0007
\mathchardef\Phi="0008
\mathchardef\Psi="0009
\mathchardef\Omega="000A
\mathchardef\varGamma="0100
\mathchardef\varDelta="0101
\mathchardef\varTheta="0102
\mathchardef\varLambda="0103
\mathchardef\varXi="0104
\mathchardef\varPi="0105
\mathchardef\varSigma="0106
\mathchardef\varUpsilon="0107
\mathchardef\varPhi="0108
\mathchardef\varPsi="0109
\mathchardef\varOmega="010A
\font\dummyft@=dummy
\fontdimen1 \dummyft@=\z@
\fontdimen2 \dummyft@=\z@
\fontdimen3 \dummyft@=\z@
\fontdimen4 \dummyft@=\z@
\fontdimen5 \dummyft@=\z@
\fontdimen6 \dummyft@=\z@
\fontdimen7 \dummyft@=\z@
\fontdimen8 \dummyft@=\z@
\fontdimen9 \dummyft@=\z@
\fontdimen10 \dummyft@=\z@
\fontdimen11 \dummyft@=\z@
\fontdimen12 \dummyft@=\z@
\fontdimen13 \dummyft@=\z@
\fontdimen14 \dummyft@=\z@
\fontdimen15 \dummyft@=\z@
\fontdimen16 \dummyft@=\z@
\fontdimen17 \dummyft@=\z@
\fontdimen18 \dummyft@=\z@
\fontdimen19 \dummyft@=\z@
\fontdimen20 \dummyft@=\z@
\fontdimen21 \dummyft@=\z@
\fontdimen22 \dummyft@=\z@
\def\fontlist@{\\{\tenrm}\\{\sevenrm}\\{\fiverm}\\{\teni}\\{\seveni}%
 \\{\fivei}\\{\tensy}\\{\sevensy}\\{\fivesy}\\{\tenex}\\{\tenbf}\\{\sevenbf}%
 \\{\fivebf}\\{\tensl}\\{\tenit}\\{\tensmc}}
\def\dodummy@{{\def\\##1{\global\let##1=\dummyft@}\fontlist@}}
\newif\ifsyntax@
\newcount\countxviii@
\def\newtoks@{\alloc@5\toks\toksdef\@cclvi}
\def\nopages@{\output={\setbox\z@=\box\@cclv \deadcycles=\z@}\newtoks@\output}
\def\syntax{\syntax@true\dodummy@\countxviii@=\count18
\loop \ifnum\countxviii@ > \z@ \textfont\countxviii@=\dummyft@
   \scriptfont\countxviii@=\dummyft@ \scriptscriptfont\countxviii@=\dummyft@
     \advance\countxviii@ by-\@ne\repeat
\dummyft@\tracinglostchars=\z@
  \nopages@\frenchspacing\hbadness=\@M}
\def\magstep#1{\ifcase#1 1000\or
 1200\or 1440\or 1728\or 2074\or 2488\or
 \errmessage{\string\magstep\space only works up to 5}\fi\relax}
{\lccode`\2=`\p \lccode`\3=`\t
 \lowercase{\gdef\tru@#123{#1truept}}}

\def\scaletype#1{\mag=#1\relax
 \hsize=\expandafter\tru@\the\hsize
 \vsize=\expandafter\tru@\the\vsize
 \dimen\footins=\expandafter\tru@\the\dimen\footins}

\def\scalefont#1#2\andcallit#3{\edef\font@{\the\font}#1\font#3=
  \fontname\font\space scaled #2\relax\font@}
\def\Mag@#1#2{\ifdim#1<1pt\multiply#1 #2\relax\divide#1 1000 \else
  \ifdim#1<10pt\divide#1 10 \multiply#1 #2\relax\divide#1 100\else
  \divide#1 100 \multiply#1 #2\relax\divide#1 10 \fi\fi}
\def\scalelinespacing#1{\Mag@\baselineskip{#1}\Mag@\lineskip{#1}%
  \Mag@\lineskiplimit{#1}}
\def\wlog#1{\immediate\write-1{#1}}
\catcode`\@=\active



\font\mybbs=msbm8
\def\bbs#1{\hbox{\mybbs#1}}

\scaletype{\magstep1}
\vsize=23.5truecm
\hsize=16.5truecm
\baselineskip=14pt
\hfuzz=1pt
\TagsOnRight

\font\kg=cmr7
\font\Mittel=cmr10 scaled \magstep 2
\font\mittel=cmr10 scaled \magstep 1
\parindent 0 pt
\def\abs{\par \hskip 20pt}

\def\pd#1#2{\frac{\partial #1}{\partial #2} }

\def\ref#1{$^{[#1]}$}
\def\sfrac#1#2{\textstyle{ \frac{#1}{#2} } \displaystyle}

\def\sgn{\hbox{sgn}}
\def\litem{\par\noindent\hangindent=\parindent\ltextindent}

\def\ltextindent#1{\hbox to \hangindent{#1\hss}\ignorespaces}
\def\chapter#1{\leftline{{\mittel{\bf #1}}}}
\def\Box#1{\mkern0.5\thinmuskip
                   \vbox{\hrule
                         \hbox{\vrule
                               \hskip#1
                               \vrule height#1 width 0pt
                               \vrule}%
                         \hrule}%
                   \mkern0.5\thinmuskip}

\def\lsbrack{\hbox{{\kg (}}}
\def\rsbrack{\hbox{{\kg )}}}
\def\n*p{\lsbrack n^{*} \! \cdot p \rsbrack}

\def\pb#1#2{ \left \{ #1, \, #2 \right \} }

\def\e{\hbox{e}}

\def\intl{\int_{-L}^{L}}

\def\lsbrack{\hbox{{\kg (}}}

\def\section#1#2{{\parindent=6truemm    \litem{#1}    #2   \par}}

\def\intm{\int_M d^2 x}

{\nopagenumbers
\line{\hss {\Mittel TPR 93-3, hep-th/9311108}}
\vskip 1truecm
\centerline{\Mittel  Light Front Quantisation}
\centerline{\Mittel  as an Initial-Boundary Value Problem}

\vskip 2 truecm

\centerline{T.~Heinzl                   \footnote"$^{1}$"{e-mail:
heinzl\@vax1.rz.uni-regensburg.d400.de}        and       E.~Werner
\footnote"$^{2}$"{e-mail:  werner\@vax1.rz.uni-regensburg.d400.de}
}
\centerline{ Institut f\"ur Theoretische Physik }
\centerline{      Universit\"at      Regensburg}
\centerline{ Universit\"atsstra\ss e 31}
\centerline{ 93053 Regensburg, FRG}

\vskip 1 truecm

$${\hbox{\hsize=14.5 truecm{\vbox{
{\bf Abstract:}\par
In the light front quantisation  scheme  initial  conditions  are
usually provided on a single lightlike hyperplane. This, however,
is  insufficient   to  yield  a  unique  solution  of  the  field
equations.  We investigate  under which additional conditions the
problem  of solving the field equations  becomes well posed.  The
consequences  for quantisation  are studied  within a Hamiltonian
formulation by using the method of Faddeev and Jackiw for dealing
with first-order Lagrangians.   For the prototype field theory of
massive  scalar  fields  in 1+1 dimensions,  we find that initial
conditions   for  fixed  light   cone  time  {\sl  and}  boundary
conditions  in the spatial  variable  are sufficient  to yield  a
consistent  commutator  algebra.   Data  on  a  second  lightlike
hyperplane  are  not necessary.   Hamiltonian  and Euler-Lagrange
equations  of motion become  equivalent;  the description  of the
dynamics remains canonical and simple. In this way we justify the
approach of discretised light cone quantisation.}}}}$$

\vskip 1truecm

\leftline{{\bf PACS:} 03.50 Kk; 11.10.Ef, Qr}

\vfill
\eject

}
\pageno=2
\leftline{\bf 1. Introduction}

\vskip 1truecm

In 1949 Dirac pointed out that in a relativistic  quantum  theory
the choice  of the time  variable  is not unique\ref{1}.   In his
"front  form" of dynamics  he suggested  to use light fronts $x^+
\equiv x^0 + x^3 = const$ as surfaces of equal "time".  This idea
was revived  about twenty  years later in the context  of current
algebra\ref{2}  as well  as deep  inelastic  scattering  and  the
parton  model\ref{3,4}  when people quantised  field theories  on
lightlike  hyperplanes  and  derived  the  corresponding  Feynman
rules\ref{5,6}.   In the recent years interest has grown in using
light    front    techniques     to    solve    field    theories
non-perturbatively.  This has been pioneered by Brodsky and Pauli
within   their   program   of   discretised   light   cone   (LC)
quantisation\ref{7},  which was later accompanied  by light front
Tamm-Dancoff  methods\ref{8}.   These  works,  however,  did  not
address a long-standing  problem of LC physics:   how can effects
related  to a non-trivial  vacuum  structure  such as spontaneous
symmetry  breaking arise from the LC vacuum which was believed to
be trivial\ref{9-11}?   This has only recently been clarified for
some model field theories by demonstrating  the important role of
modes  with  LC momentum  $k^+  =0$ which  carry  the non-trivial
vacuum aspects\ref{12,13}.   In all of these approaches one makes
explicit or implicit assumptions  on the behaviour  of the fields
for large values  of the spatial  variable  $x^- \equiv x^0 - x^3
\to  \pm  L \to \pm \infty$  in terms  of asymptotic  or boundary
conditions  (BC).   It  has  never  been  checked  whether  these
conditions  which  must  hold  for {\sl all} LC times  $x^+$  are
consistent with the dynamics of the system. It has, however, been
suggested  that  the  system  might  so become   over-determined.
Already in the early seventies it was noted that quantising  on a
light front means quantising  on the characteristic  surfaces  of
the (classical)  field equations\ref{14,15}.   The conclusion was
that one is dealing with a characteristic  initial  value problem
(CIVP) when one wants to solve these equations. This implies that
one has to specify  data  on both characteristics  $x^+ = const$,
$x^- = const$.  Whether this is of any relevance  for the quantum
theory  has  not  been  studied  before  McCartor's  work  on  LC
quantisation of massless fields\ref{16}. He found that one should
quantise  free  massless  fermions  in  1+1  dimensions  on  both
characteristics,   which    essentially    means   that  one  has
equal-$x^+$  and  equal-$x^-$  commutators.    In  a  Hamiltonian
picture, this amounts to having two different times. Accordingly,
the  Poincar\'e  generators   receive  contributions   from  both
characteristics.    Though  this  method  is  surely  legitimate,
quantising on a cone-like surface instead of a single light front
makes  it  difficult  to  establish  a  Hamiltonian  formulation.
Therefore,  in this  work  we address  the  alternative  question
whether  it is possible  to choose  a single  light  front $x^+ =
const$  as  the  initial   surface   and  develop   a  consistent
Hamiltonian  formalism which describes the dynamics completely in
the parameter $x^+$. We concentrate on the case of massive scalar
fields  in 1+1 dimensions,  which  turns  out  to be sufficiently
general.  Only a few remarks will be made concerning the massless
case, which is generically different. \abs

\vfill\eject

Our starting  point is to clarify  the relation  between  initial
data for the Klein-Gordon (KG) equation and the specification  of
canonical  commutators  for  the  cases  of  both  spacelike  and
lightlike hypersurfaces.   We find that the independent canonical
commutators  can be viewed as special  initial data satisfying  a
consistency  condition  which serves as a testing  ground for the
quantisation prescription.  These results are essentially derived
by considering the Euler-Lagrange  (EL) equation of motion.  As a
next  step  we  perform  an  analogous  investigation   within  a
Hamiltonian  picture.  In former publications  we have set up the
Hamiltonian formulation\ref{13}  by treating LC field theories as
constrained    systems    in    the    sense    of   Dirac    and
Bergmann\ref{17,18}.   The outcome  of this procedure  were Dirac
brackets,  i.e.~modifications    of   the  conventional   Poisson
brackets, which became the commutators after quantisation.   This
has been called a generalised  correspondence  principle.  It has
turned out recently,  however, that there is a much more economic
way of deriving the basic (or Dirac) brackets,  namely the method
of Hamiltonian reduction due to Faddeev and Jackiw\ref{19}.  This
method  is also  more intuitive;  for the case at hand  it simply
reduces to demanding equivalence of EL and Hamiltonian  equations
of motion.   This already  determines  the basic  bracket.   More
technically,  within the new method Dirac's  primary constraints,
stemming  from  the  definition  of the  canonical  momenta,  are
completely absent.\abs
For our investigation the benefit of the Faddeev-Jackiw procedure
is a very transparent analysis   of  the sensitivity of the basic
field brackets (or commutators)  on the chosen initial and/or BC.
It turns out that characteristic data specified on the {\sl cone}
$x^\pm = const$ together with the canonical LC Hamiltonian do not
lead to a consistent  description  of the dynamics.   There is no
canonical field commutator such that the Hamiltonian  equation of
motion  matches  the  EL  one  (i.e.~the KG equation).   The only
consistent  way  of  quantising  on a single  light  front  is to
replace the data on the second characteristic  by appropriate  BC
that hold for all LC times.  The dynamics is then well determined
by the Hamiltonian  and in accordance  with the general (Riemann)
solution of the KG equation.   This we prove quite generally  and
check    explicitly     for    the    Pauli-Jordan\ref{20}     or
Schwinger\ref{21} function which is the free field commutator for
arbitrary LC time differences.   The essence of the proof is that
the data on the second characteristic, $x^- = const$, are already
determined  by the data on $x^+  = const$  {\sl and} the BC.  The
former, therefore, can no longer be specified independently,  or,
put in the quantum  language:   a single  equal-$x^+$  commutator
together with suitable BC in $x^-$ is sufficient. In view of this
result, which mathematically  corresponds  to an initial-boundary
value problem, we regard  the term "light front quantisation"  as
more appropriate than "light cone quantisation".\abs
The paper is organised  as follows.   In Section  2 we review the
relation  between  instantaneous  quantisation  (on  a  spacelike
hypersurface)  and the Cauchy  problem  for the KG equation.   In
Section 3 we compare quantisation  on lightlike  hyperplanes  and
the CIVP.  The Hamiltonian  formulation  is set up in Section  4,
whereas, in Section 5, we analyse the relation between (periodic)
BC and  the solution  of the  CIVP.   Finally,  in Section  6, we
discuss some consequences of this work.

\vfill\eject

\leftline{\bf 2. Instantaneous Quantisation as a Cauchy Problem}

\vskip 1truecm

We start from the covariant action

$$  S  =  \intm  \,  {\cal  L}  = \intm  \,  \left[  \sfrac{1}{2}
\partial_{\mu}  \phi  \, \partial^{\mu}  \phi - \sfrac{1}{2}  m^2
\phi^2 \right] \; , \tag 2.1 $$

where $M$ denotes Minkowski space. The priniciple of least action
yields

$$ \delta  S = \intm  \, \left[\pd{\cal  L}{\phi}  - \partial_\mu
\frac{\partial  {\cal  L}}{\partial  (\partial_{\mu}\phi)}\right]
\delta  \phi  + \int_{\partial  M}  d\sigma_{\mu}  \frac{\partial
{\cal  L}}{\partial  (\partial_{\mu}\phi)}  \delta  \phi = 0 \; .
\tag 2.2 $$

If  we do not  vary  on the  boundary  $\partial  M$ of Minkowski
space, $\delta \phi \vert_{\partial  M} = 0$, the surface term in
$\delta S$ vanishes and we obtain the (massive) KG equation

$$  (\Box{2.5truemm} + m^2) \phi = 0  \; . \tag 2.3  $$

This  is a hyperbolic  partial  differential  equation  (PDE)  of
second  order\ref{22,23}.  The  Cauchy  problem  for  it  can  be
formulated as follows:  determine a solution $\phi (x^0, x^1)$ of
(2.3) which satisfies the initial conditions

$$ \phi(x^0 = 0 , x^1) = f(x^1)\; ; \quad \dot \phi(x^0 = 0, x^1)
= g(x^1) \; . \tag 2.4 $$

The prescribed functions, $f$ and $g$, which determine $\phi$ and
its time derivative  $\dot  \phi$  on the surface  $x^0 =0$,  are
called  Cauchy data.  The Cauchy problem  for the KG equation  is
well posed, i.e.~it  has a unique solution  which we are going to
derive now. As $\phi$ obeys (2.3) its Fourier representation must
be of the form

$$ \phi (x) = \int \frac{d^2 p}{2\pi} \delta(p^2  - m^2) \chi (p)
\e^{-i p\cdot x} \; . \tag 2.5  $$

After  performing  the $p^0$-integration  and an inverse  Fourier
transform we obtain the mass-shell values of $\chi$,

$$ \chi(p^0  = \pm \omega_p,  p^1)  = \int  dx^1  \e^{-ip^1  x^1}
[\omega_p  f (x^1) \pm g(x^1) ] \equiv
\omega_p f(p^1) \pm g(p^1)\; , \tag 2.6 $$

where  $\omega_p  = (p_1^2  + m^2)^{1/2}$.   On  the  mass-shell,
therefore,  $\chi$  is uniquely  determined  by the Cauchy  data.
After quantisation the amplitudes $\chi$ get replaced by creation
and annihilation  operators  $a$ and $a^\dagger$;  equation (2.6)
then expresses the well known fact that these Fock operators  are
given  in  terms  of  the  field  {\sl  and}  the  velocity.   By
reinserting  (2.6)  into (2.5) the field $\phi$  can be expressed
through its initial values, $f$ and $g$,

$$ \phi (x) = \int dy^1 \left[f(y^1) \pd{}{y^0} \Delta (x-y) -
g (y^1) \Delta (x-y) \right] \; , \tag 2.7  $$

where   we   have   introduced   the   Pauli-Jordan\ref{20}    or
Schwinger\ref{21} function (or better distribution)

$$ \Delta (x) = -\frac{i}{2\pi}  \int d^2 p \,\delta  (p^2 - m^2)
\sgn(p^0)  \e^{-i  p \cdot  x} = -\frac{1}{2}  \sgn (x^0)  \theta
(x^2) J_0 (m \sqrt{x^2}) \; . \tag 2.8 $$

It  is  an  invariant  solution  of  the  KG equation,  which  is
antisymmetric, $\Delta (x) = - \Delta (-x)$, and vanishes outside
the LC,

$$ \Delta (x) = 0 \; , \quad \hbox{for}  \quad x^2 < 0 \; .  \tag
2.9 $$

These properties,  however,  are just the requirements  which are
to be satisfied  by the commutator  of two free scalar fields  so
that one can identify

$$  [\phi (x), \phi (0)] = i \Delta (x) \; , \tag 2.10  $$

where the factor $i$ is necessary for hermiticity.  An exhaustive
discussion   of  these   issues   can  be  found   in  Schweber's
textbook\ref{24}.    With  the  help  of  (2.8)  one  infers  the
equal-time commutators

$$\eqalignno{[\phi   (x),  \phi  (0)]_{x^0   =  0}  &=  i  \Delta
(x)\vert_{x^0 = 0} = 0 \; , & (2.11a) \cr
[\dot \phi   (x),  \phi  (0)]_{x^0   =  0}  &=  i \dot \Delta
(x)\vert_{x^0 = 0} = -i \delta (x^1) \; . & (2.11b) \cr}  $$

This is completely  consistent  with the general  formula  (2.7).
Replacing there $\phi$, $f$ and $g$ by the commutators,

$$ [\phi  (x), \phi (0)] = \int dy^1 \left\{[\phi  (y), \phi (0)]
\pd{}{y^0}  \Delta (x-y) - [\dot \phi (y), \phi(0)]  \Delta (x-y)
\right\}_{y^0 = 0} \; , \tag 2.12 $$

and inserting  (2.11) gives just the identity  (2.10).   Equation
(2.12) is therefore a consistency  condition which expresses  the
commutator  for arbitrary time differences  $x^0 > 0$ through its
Cauchy  data  at time  $x^0  =0$.   The latter  are just  the two
independent canonical equal time commutators (2.11).  To quantise
consistently {\sl both} of them have to be specified.\abs
In the next sections we want to analyse the relation between data
and commutators  if one chooses a light front as the quantisation
hypersurface.   To  this  end  we  will  derive  the  appropriate
consistency condition analogous to (2.12).

\vfill\eject

\leftline{\bf  3.  Light  Cone  Quantisation  as a Characteristic
Initial Value Problem}

\vskip 1truecm

Let us rewrite  the KG equation  (2.3)  in terms  of LC variables
$x^\pm$, with $\partial^\pm = 2\partial/\partial x^\mp$,

$$ (\partial^+ \partial^- + m^2) \phi = 0 \; . \tag 3.1 $$

The  lightlike  hyperplanes   $x^\pm  =  const$  are  called  the
characteristics of this hyperbolic PDE (cf.~any textbook on PDEs,
e.g.~Ref.s  [22,23]).   It has been shown explicitly  in a recent
preprint\ref{25}   that  specifying   the  field  $\phi$  and  an
arbitrary number of derivatives  on one characteristic  only does
not lead to a unique  solution  of (3.1).   This is not the case,
however,  for the CIVP which is defined as follows:  Determine  a
solution $\phi (x^+, x^-)$ which satisfies the initial conditions

$$  \eqalignno{\phi (x^+, x_0^- ) &= f (x^+) \; , & (3.2a) \cr
\phi (x_0^+, x^-) &= g(x^-) \; ,  & (3.2b) \cr}  $$

and the continuity condition

$$ \phi (x_0^+ , x_0^-) = f(x_0^+) = g(x_0^-) \; . \tag 3.3 $$

The functions  $f$ and $g$ which  specify  $\phi$  on {\sl  both}
characteristics  are called  characteristic  data.  According  to
Neville  and  Rohrlich\ref{14}  the solution  of the CIVP  can be
obtained from the solution (2.7) of the Cauchy problem via Gauss'
theorem. To this end one introduces the quantity\ref{14,24}

$$ F_\mu (x,y) = \Delta  (x-y)  \pd{}{y^\mu}  \phi (y) - \phi (y)
\pd{}{y^\mu} \Delta (x-y) \; , \tag 3.4  $$

which is divergence free,

$$ \pd{}{y^\mu} F^\mu (x,y) = 0 \; , \tag 3.5  $$

since  both $\phi$  and $\Delta$  satisfy  the KG equation.   One
integrates  (3.5) over the volume  bounded  by the simplex  $ABC$
(Fig.1), where on $AB$:  $x^0 =0$, on $BC$: $x^- = x_0^-$  and on
$CA$: $x^+ = x_0^+$. Gauss' theorem yields

$$ 0= (\int_{AB} + \int_{BC} + \int_{CA}) d\sigma (y) n_\mu F^\mu
(x,y) \; , \tag 3.6  $$

with $d\sigma$  and $n_\mu$ the appropriate  surface elements  and
normal unit vectors. Explicitly one has

$$ \eqalign{&\int_{AB} dy^1 \left[ \phi(y) \pd{}{y^0} \Delta (x-y)
- \Delta (x-y) \pd{}{y^0} \phi (y) \right] = \cr
= &\int_{CB} dy^+ \left[ \phi(y) \pd{}{y^+} \Delta(x - y) - \Delta
(x - y) \pd{}{y^+}  \phi  (y)  \right]  + \cr
+ &\int_{CA}  dy^-  \left[  \phi(y)  \pd{}{y^-}  \Delta(x  - y) -
\Delta (x - y) \pd{}{y^-}\phi (y) \right] \; . \cr} \tag 3.7 $$

According   to  (2.9),  $\Delta  (x-y)$  vanishes  for  spacelike
argument,  $(x-y)^2  < 0$, and one can extend the integration  on
the l.h.s.~to  infinity.  Comparison  with (2.7) then reveals the
l.h.s.~as  the solution of the Cauchy problem so that one finally
has with the data (3.2)

$$ \eqalign{\phi (x^+, x^-) &= \int_{x_0^+}^\infty  dy^+ \left[
f(y^+) \pd{}{y^+} \Delta(x^+  - y^+, x^- - x_0^-) - \Delta (x^+ -
y^+, x^- - x_0^-) \pd{f}{y^+}  \right] + \cr
&+ \int_{x_0^-}^\infty dy^- \left[ g(y^-) \pd{}{y^-} \Delta(x^+ -
x_0^+,  x^- - y^-) - \Delta  (x^+ - x_0^+, x^- - y^-) \pd{g}{y^-}
\right] \; . \cr} \tag 3.8 $$

This is the desired solution  of the CIVP for the KG equation  in
terms  of the initial  data,  $f$ and $g$,  and the  Pauli-Jordan
function $\Delta$, which in LC coordinates can be written as

$$ \Delta (x^+ , x^- ) = - \sfrac{1}{4} [\sgn (x^+) + \sgn (x^-)]
J_0 (m\sqrt{x^+ x^-}) \; . \tag 3.9 $$

The  Bessel  function  $J_0$  is the Riemann  function  of the KG
equation,  and for $x^\pm \ge x_0^\pm$  equation  (3.8) coincides
with the classical  Riemann solution for hyperbolic  PDEs applied
to  the  case  at  hand   (often   also   called   the  telegraph
equation)\ref{22,23},

$$ \eqalign{\phi  (x^+ , x^-) &= \sfrac{1}{2}  \bigl [f(x_0^+)  +
g(x_0^-)\bigr] R(x^+ - x_0^+, x^- - x_0^-) + \cr
&+ \sfrac{1}{2} \int_{x_0^-}^{x^-} dy^- R(x^+ - x_0^+, x^- - y^-)
\partial_y^+ g (y^-) + \cr
&+ \sfrac{1}{2} \int_{x_0^+}^{x^+} dy^+ R(x^+
- y^+, x^- - x_0^-) \partial_y^- f(y^+) \; . \cr}  \tag 3.10  $$

The Riemann function $R$ is explicitly given by

$$ R(x^+,  x^-)  = J_0  (m\sqrt{x^+  x^-})  = \sum_{n=  0}^\infty
\frac{1}{n! \, n!}(-\sfrac{1}{4} m^2 x^+x^-)^n  \; . \tag 3.11 $$

For practical  calculations  it is often more convenient  to work
with the Riemann solution  because  one is dealing with functions
instead of distributions.\abs
As a side remark  we note  that  the solutions  (3.8)  and (3.10)
behave  smoothly  in the limit  of vanishing  mass,  which  will,
however, be discussed elsewhere.\abs
We emphasize that, in order to obtain a unique solution,  data on
{\sl both} characteristics  and not only on a single  light front
have to be specified.   For the quantum theory (in $d=1+1$)  this
implies quantisation on the light {\sl cone}: the two independent
commutators that have to be specified are

$$ [\phi (x), \phi (0)]_{x^\pm = 0} = i \Delta (x)\vert_{x^\pm  =
0} = - \sfrac{i}{4} \sgn (x^\mp) \; . \tag 3.12  $$

This is obtained  from (3.9) by setting

$$  \sgn (x^\pm  = 0) = 0 \; , \tag 3.13  $$

which  can  be  verified  from  the  Fourier  representation   of
$\Delta$\ref{14}.    Let  us  derive  the  consistency  condition
analogous to (2.12).  To this end we set $x_0^\pm = 0$, $f(x^+) =
-\sfrac{i}{4}  \sgn (x^+)$, $g(x^-) = -\sfrac{i}{4}  \sgn (x^-)$.
Then,  after a partial  integration  in (3.8),  $[\phi  (x), \phi
(0)]$ should be given by

$$ \eqalign{[\phi  (x), \phi (0)] & {\buildrel ?  \over =} \left.
-\sfrac{i}{4} \sgn(y^+) \Delta(x^+ - y^+, x^-)\right\vert_{y^+  =
0}^\infty + i \int_0^\infty  dy^+ \delta (y^+) \Delta (x^+ - y^+,
x^-) + \cr
&+\left.   -\sfrac{i}{4}
\sgn(y^-) \Delta(x^+ , x^- - y^-)\right\vert_{y^-  = 0}^\infty + i
\int_0^\infty  dy^- \delta  (y^-)  \Delta  (x^+ , x^- - y^-) \; .
\cr} \tag 3.14 $$

The surface  terms vanish due to (2.9) and (3.13);  the integrals
give $\sfrac{i}{2}  \Delta (x)$ each which, shows that (3.14)  is
really an identity. The consistency condition is then

$$ \eqalign{[\phi (x), \phi (0)] &= \int_{0}^\infty  dy^+ \left\{
[\phi (y) , \phi (0)] \pd{}{y^+}  \Delta(x  - y) - \Delta (x - y)
\pd{}{y^+}  [\phi (y), \phi (0)] \right\}_{y^- = 0} + \cr
&+ \int_{0}^\infty  dy^- \left\{
[\phi (y) , \phi (0)] \pd{}{y^-}  \Delta(x  - y) - \Delta (x - y)
\pd{}{y^-} [\phi (y), \phi (0)] \right\}_{y^+ = 0} \; . \cr} \tag
3.15  $$

It expresses  the  commutator  for all  $x^\pm$  in terms  of its
characteristic  values (3.12).  Again this tells us that the CIVP
corresponds to quantisation on the cone.  For the quantisation of
massless  scalar  fields in $d = 1+1$ the commutators  (3.12) are
postulated in the textbook Ref.~[26]. McCartor has used analogous
expressions   for  massless  fermions\ref{16}.    In  two  recent
preprints\ref{27}  quantisation  on both $x^+ = const$ and $x^- =
const$  has been performed  for Yukawa theory  and abelian  gauge
fields (in $d= 3+1$). Accordingly, one has to introduce {\sl two}
time  parameters,  $x^+$  {\sl  and}  $x^-$.   In the rest of the
literature   on  "light  cone"  or  "light  front"  quantisation,
commutators are prescribed on one characteristic  (usually $x^+ =
0$)  only.   This  is  indispensable  if  one  wants  to  have  a
Hamiltonian formulation with {\sl one} single evolution parameter
$x^+$, but seems to be in contradiction  with the results  above.
\abs
Before we answer the question,  whether quantisation  on a single
light  front  is possible  at all,  we want  to analyse  why  the
relevance of data on a second characteristic (even in the massive
case)  has  been  overlooked  so far.   For this  purpose  let us
consider the Fourier representation (2.5) of $\phi$ written in LC
variables,

$$ \phi (x^+ , x^-) = \frac{1}{4\pi}  \int dp^+ dp^- \e^{-ip^+
x^-/2 - i p^- x^+ /2} \delta(p^+  p^- - m^2) \chi (p^+, p^-) \; .
\tag 3.16  $$

The  (erroneous)  claim  in the literature  is now that  the Fock
operators   which  are  obtained  from  quantising   the  Fourier
amplitude $\chi$ can be expressed through the field $\phi$ alone,
specified  on  one  characteristic\ref{6,9,11}.  This  should  be
contrasted  with (2.6) where both the field  and the velocity  on
the initial surface are needed.   Mathematically, the above claim
amounts to assuming the identity

$$ \delta (p^+ p^- - m^2) = \frac{1}{\vert p^+ \vert} \delta (p^-
- m^2 / p^+) \; , \tag 3.17  $$

which leads to

$$ \int dx^- \e^{ip^+  x^- /2} \phi (x^+ , x^- ) \equiv \phi (x^+
,  p^+)  = \frac{1}{\vert  p^+  \vert}  \chi  (p^+  , m^2  / p^+)
\e^{-im^2 x^+ \! /2p^+} \; ,  \tag 3.18  $$

and (if we set $x^+ = 0$)

$$ \chi (p^+ , m^2 /p^+)  = \vert  p^+ \vert \, \phi (x^+ = 0, p^+)
\equiv \vert p^+ \vert \,  g(p^+) \; . \tag 3.19 $$

So  it  really  seems  that  the  amplitude   $\chi$  is  totally
determined  by the field $\phi$ on the characteristic  $x^+ = 0$,
the Fourier  transform  of which  is $g(p^+)$.   The data  on the
second  characteristic, which  are  represented  by the  function
$f(p^-)$,  do not enter  at all.   This  surely  contradicts  the
solutions  (3.8)  and (3.10)  which  depend  on both data.   Note
however, that the equations (3.17) and (3.18) are ill-defined  at
$p^+ = 0$.  Both are actually identities for distributions rather
than  functions, and  $1/\vert  p^+  \vert$  is not defined  as a
distribution\ref{28},   unless   one   specifies   some   further
regulating prescription. One possibility is to think of $\chi$ as
being a test function and demand\ref{29}

$$ \lim_{p^+  \to 0} \, \frac{1}{p^+}  \chi (p^+ , m^2 / p^+) = 0
\; . \tag 3.20 $$

This can be translated into a condition on $\phi$: performing the
limit $p^+ \to 0$ in (3.18)  and using the KG equation  we obtain
with (3.20)

$$ 0= \int  dx^-  \phi = -  \frac{1}{m^2}  \int  dx^-  \partial^+
\partial^-  \phi  = - \frac{2}{m^2}  \left[\partial^-  \phi (x^+,
\infty)  - \partial^-  \phi (x^+, -\infty)  \right]  \; . \tag
3.21 $$

We see that the postulate (3.20) is equivalent  to demanding that
$\partial^- \phi$ and therefore $\phi$ be {\sl periodic} in $x^-$
for all $x^+ \ge 0$.  Clearly,  this is a severe  restriction  on
$\phi$,  and  one  has to verify,  whether  such  a condition  is
consistent  with the characteristic  data (3.2)  and the solutions
(3.8) and (3.10).   This will be done in the next section when we
study the Hamiltonian formulation. \par
The moral  so far is that  there  is a subtle  interplay  between
initial   or  boundary   data  with  respect  to  $x^-$  and  the
singularity  in the conjugate  variable, i.e.~at $p^+ = 0$.  Only
with sufficient conditions guaranteeing a well-defined inverse of
$p^+$  do  the  manipulations   leading  to  (3.19)  make  sense,
i.e.~only  if $\phi$ is periodic in $x^-$, the statement that the
data $f(x^+)$  are not needed seems to be correct.   This will be
verified  in more detail in Section  5.  It will also become more
evident in the following  that inverting  the variable  or better
distribution   $p^+$   is  {\sl  the}  problem   of  light  front
quantisation.

\vskip 2truecm

\leftline{\bf 4. The Hamiltonian Formulation}

\vskip 1truecm

In this section we want to analyse the relation  between the data
for the KG equation (viewed as a Hamiltonian  equation of motion)
and the basic bracket  between  the field variables,  which is to
become   the   commutator    upon   quantisation.     In   former
publications\ref{13}  we obtained this bracket as a Dirac bracket
resulting  from  the  Dirac-Bergmann   algorithm\ref{17,18}   for
dealing  with constrained  systems.   As the kinetic  term in the
Lagrangian  (2.1), $\partial^+  \!  \phi \partial^-  \!\phi$,  is
linear in the LC velocity, $\partial^-  \phi \equiv 2 \dot \phi$,
the Lagrangian is called singular\ref{18}.   This observation was
then the starting point for running the Dirac-Bergmann machinery.
However,  as has been noted  by Faddeev  and Jackiw\ref{19},  the
appearance  of (primary) constraints  for first order Lagrangians
is an artifact  of the method  one uses rather  then  of physical
significance.   To avoid  the  complexity  of the  Dirac-Bergmann
algorithm they developed a much more economic method which in its
simplest  form  reduces  to  demanding  equivalence   of  EL  and
Hamiltonian   equations.    This  already  determines  the  basic
brackets\footnote"$^1$"{The  method  of Faddeev  and  Jackiw  has
recently  been  applied  to  the  quantisation   of  Light  Front
QCD\ref{30}.  There, the basic brackets have been obtained as the
symplectic  structure of the (infinite-dimensional)  phase space.
This essentially boils down to the way we do it\ref{19}.}.\abs
As the Hamiltonian  equations express the velocity $\dot \phi$ in
terms of the fields, it is necessary to perform  a first integral
of  the  KG equation,  which  amounts  to invert  the  derivative
$\partial^+$  by choosing an appropriate  Green function $G(x^- -
y^-)$.   In momentum  space, this inverse  is a properly  defined
distribution  $G(p^+)$ replacing the naive expression  $1/p^+$ as
was mentioned above. One finds

$$ \dot \phi (x^+, x^-) = -\frac{m^2}{4}  \int dy^- G (x^-,  y^-)
\phi (x^+ , y^-) + h (x^+) \; . \tag 4.1 $$

Up to possible boundary or finite volume terms the Green function
$G$ has to obey

$$ \frac{d}{dx^-}  G (x^- , y^- ) = \delta (x^- - y^-) \; ,  \tag
4.2  $$

or in momentum space

$$ \sfrac{1}{2} p^+ G( p^+) = 1 \; . \tag 4.3 $$

Clearly,  (4.2) and (4.3) do not determine  $G$ uniquely which is
partially  reflected  by the appearance  of the  function  $h$ in
(4.1).  Both $G$ and $h$ have to be determined  by initial  and/or
BC, which will be done below. In order to demonstrate the way how
the  method  of Faddeev  and Jackiw  works,  however,  the formal
expression (4.1) is sufficient. \abs
The canonical hamiltonian of the KG system is

$$ H_c \equiv  \sfrac{1}{2}  P^-  = \sfrac{1}{4}  \int  dx^-  m^2
\phi^2 \; , \tag 4.4 $$

so that the Hamiltonian equation of motion becomes

$$ \eqalign{\dot  \phi (x^+ , x^-) &= \pb{\phi (x^+, x^-)}{H_c}^*
+ \pd{\phi}{x^+}(x^+, x^-) = \cr
&=  \frac{m^2}{4}  \int dy^- \pb{\phi  (x^+, x^-)}{\phi^2(x^+  ,
y^-)}^* + \pd{\phi}{x^+}(x^+, x^-) \; . \cr}  \tag 4.5 $$

Here    $\pb{\;}{}^*$    denotes    the   basic    bracket    and
$\partial\phi/\partial  x^+$ a possible  explicit time dependence
of $\phi$  which  is not determined  by the Hamiltonian.   As any
reasonable bracket should obey a Leibniz rule, one gets

$$ \dot  \phi  (x^+ , x^-) =  \frac{m^2}{2}  \int  dy^- \pb{\phi
(x^+,   x^-)}{\phi(x^+   ,   y^-)}^*   \phi   (x^+   ,   y^-)   +
\pd{\phi}{x^+}(x^+, x^-) \; . \tag 4.6 $$

This coincides with (4.1) if one identifies

$$ \eqalignno{G  (x^- , y^-) &\equiv - 2 \pb{\phi  (x^+, x^-)}{\phi
(x^+ , y^-)}^* \; , & (4.7a) \cr
h(x^+) &\equiv \pd{\phi}{x^+} (x^+, x^-) \; . & (4.7b) \cr} $$

One notices that the basic bracket  of two fields $\phi$ is given
entirely in terms of the Green function  $G$, i.e.~the inverse of
the derivative  $\partial^+$.   To emphasize  it once more,  this
inverse is only well defined if one specifies sufficient  initial
and/or  boundary  data.   If this has been done, quantisation  is
straightforward  (up to possible operator ordering problems), but
then, of course  dependent  on the prescribed  data.\abs
Two  properties   the  Green  function  $G$  should  fulfill  are
immediately  clear  if  it  is  to  become  a commutator,  namely
translation invariance and antisymmetry,

$$  \eqalignno{G (x^-, y^- ) &= G (x^- - y^-) \; , & (4.8a) \cr
G (x^- -y^-) &= - G(y^- - x^-) \; . & (4.8b)  \cr}  $$

In the following  we will discuss three examples.  In all of them
we specify  data on the light  front $x^+ =0$, but the conditions
dependent on $x^-$ will be different.\abs
Let us start with the characteristic  data (3.2) with $x_0^\pm  =
0$.  Using the following identity for the Riemann function $R$ of
(3.11),

$$ \pd{}{x^+} R(x^+ , x^- ) = - \frac{m^2}{4}  \int_0^{x^-}  dy^-
R(x^+ , y^-) \; , \tag 4.9  $$

one finds that the Riemann solution (3.10) obeys

$$ \dot \phi (x^+, x^-) = -\frac{m^2}{4} \int_{0}^{x^-}  dy^-
\phi (x^+, y^-) + \dot f (x^+)  \; . \tag 4.10  $$

The dependence  on the data $f(x^+)$ on the second characteristic
is clearly exhibited.  Equation (4.10) corresponds  to the choice
of the Green function

$$ G_1 (x^-, y^-) = \theta (x^- - y^-) - \theta  (- y^-) \;
, \tag 4.11  $$

which  is vanishing  on the  second  characteristic,  $x^-  = 0$.
Obviously,  this Green function  breaks  translation  invariance.
This can be cured by setting\ref{14} $x_0^- \to -\infty$, so that
$G_1$ becomes

$$  G_1 (x^- - y^- ) = \theta (x^- - y^-) \; , \tag 4.12  $$

but still  one does  not have antisymmetry.   $G_1$  is therefore
inappropriate as a bracket or a commutator and must be abandoned.
The moral  is that quantisation  on a single  light  front  using
characteristic  data  and  the  canonical  Hamiltonian  (4.4)  is
impossible!    This  does  not  come  as  a  surprise  since  the
Hamiltonian  should be an integral over the initial value surface
and in the case at hand would imply also integration  over $x^- =
const$\ref{16}.  Recently, a Hamiltonian formulation has been set
up which uses both $x^+$ and $x^-$  as evolution  parameters  and
therefore    requires    two    Hamiltonians    generating    the
time development\ref{27}.\abs
What we want to pursue instead is to investigate whether there is
any way to use the simple Hamiltonian (4.4), quantise on a single
light front and have a consistent  formulation  for the dynamics.
Hence, from now on, we will follow the philosophy of "discretised
LC quantisation"\ref{7}  and enclose  the  physical  system  in a
finite  spatial  volume,  $ -L \le x^- \le L$.  This serves  as a
regulator  in the infrared and makes the momentum variable  $p^+$
discrete  and denumerable.   In a finite  volume,  BC have  to be
specified  and  these  will  be  of crucial  importance  for  the
following.  First note that BC with respect to $x^-$ have to hold
for {\sl all} LC times $x^+$ and can therefore  have an influence
on the dynamics. In the last section we have already seen that BC
can do a good job in making  the inverse  of $p^+$  well-defined.
Let us investigate this now in greater detail.  We begin with the
case of antiperiodic BC,

$$  \phi (x^+, -L) + \phi (x^+ , L) = 0 \; . \tag 4.13  $$

These are usually used for fermions\ref{7} but sometimes also for
bosonic fields\ref{30}. The corresponding Green function is

$$ G_2 (x^- , y^-) = \sfrac{1}{2} \sgn (x^- - y^-) \; , \tag 4.14
$$

which itself is antiperiodic,

$$  G_2(L, y^- ) + G_2 (-L, y^-) = 0 \; . \tag 4.15  $$

In momentum space this corresponds to the principal value

$$ \sfrac{1}{2} G_2(p^+) = {\cal P} \frac{1}{p^+} \; , \tag 4.16 $$

which   is  the   canonical   regularisation   of  the   function
$1/p^+$\ref{28}. According to (4.1) the time derivative of $\phi$
is given by

$$\dot  \phi (x^+,  x^-) = - \frac{m^2}{4}  \intl  dy^- G_2 (x^-,
y^-) \phi (x^+, y^-) + h_2 (x^+) \; . \tag 4.17  $$

Antiperiodicity requires

$$ h_2 (x^+) = 0 \; . \tag 4.18  $$

As $G_2$ is the unique inverse  of $\partial^+$  in this case and
fulfills  all  necessary   conditions   we  finally  obtain  from
(4.7$a$) the consistent bracket

$$  \pb{\phi(x^+,  x^-)}{\phi(x^+,  y^-)}^*  = -\sfrac{1}{4}  \sgn
(x^- - y^-)\; . \tag 4.19 $$

After quantisation this coincides with the equal-$x^+$-commutator
of   (3.12).     This    form   is   commonly    used    in   the
literature\ref{4,6,31};   the  connection  with  the  chosen  BC,
however, has only been discussed by a few authors\ref{32} and not
been entirely clarified.   \par
The discussion  above can be still slightly generalised:   Let us
consider the BC

$$ \phi (x^+, L) + \phi (x^+, -L) = 2 b(x^+) \; , \tag 4.20 $$

with an arbitrary function $b$. This yields

$$ \dot \phi (x^+, x^-) = -\frac{m^2}{4}  \intl dy^- \sfrac{1}{2}
\sgn (x^- - y^-) \phi (x^+, y^-) + \dot b(x^+) \; . \tag 4.21 $$

The field $\phi$ can be decomposed,

$$ \phi (x^+, x^-) = \varphi  (x^+, x^-) + b(x^+)  \; , \tag 4.22
$$

such  that $\varphi$  is antiperiodic.   Again,  in this case,  a
consistent quantisation  is possible but now requires an explicit
time dependence  of $\phi$, $\partial  \phi / \partial x^+ = \dot
b(x^+)$.   The  use of the BC (4.20)  offers  the possibility  of
introducing a spatially constant background field (represented by
$b$)  and  could  therefore  be an alternative  to the method  of
Ref.~[13], where periodic BC have been used.\abs
Periodic BC, however,  also turn out to be consistent,  as we are
now going to show. They read explicitly

$$  \phi (x^+, L) - \phi (x^+ , -L) = 0 \; . \tag 4.23 $$

The appropriate Green function is the periodic sign function

$$ G_3 (x^- , y^-) = \sfrac{1}{2}  \sgn (x^- - y^-) - \frac{x^- -
y^-}{2L} \; , \tag 4.24  $$

which  is antisymmetric,  translation  invariant  and  obeys  the
conditions

$$ \eqalignno{\pd{}{x^-}  G_3(x^-  - y^-) &= \delta (x^- - y^-) -
\frac{1}{2L} \; , & (4.25a) \cr
G_3(L, y^- ) - G_3 (-L, y^-) &= 0 \; . & (4.25b) \cr}  $$

This gives

$$ \dot \phi (x^+,  x^-) = -\frac{m^2}{4}  \intl  dy^- G_3 (x^- ,
y^-) \phi (x^+, y^-) + h_3 (x^+) \; . \tag 4.26  $$

In contradistinction  to the antiperiodic  case  the BC (4.25$b$)
obeyed by the Green function do {\sl not} determine  the function
$h_3$.  However,  $G_3$ obeys still another  condition:  it has a
vanishing zero mode,

$$  \intl dx^- G_3 (x^- , y^-) = 0 \; . \tag 4.27  $$

Therefore, integrating (4.26) gives rise to

$$ \frac{1}{2L} \intl dx^- \dot \phi (x^+, x^-) = h_3 (x^+) \;
. \tag 4.28  $$

By the same argument as in (3.21) we know that the integral  over
$\phi$  and  therefore  (by  differentiation)  over  $\dot  \phi$
vanishes.  Thus $h_3$ is determined to be zero. Again one finds a
consistent bracket

$$  \pb{\phi  (x^+,  x^-)}{\phi   (x^+,  y^-)}^*  =  \sfrac{1}{2}
\left[\sfrac{1}{2} \sgn (x^- - y^-) - \frac{x^- - y^-}{2L}\right]
\; . \tag 4.29 $$

This  bracket  has earlier  been obtained  via the Dirac-Bergmann
algorithm\ref{13} but much more effort was necessary.\abs
With  regard  to the  theory  of PDEs,  demanding  BC alters  the
problem which is to be solved. One does no longer have a CIVP but
rather  an initial  value  problem  in the variable  $x^+$  and a
boundary  value  problem  with  respect  to  $x^-$.  This  is  an
initial-boundary value\ref{23} or "mixed"\ref{22} problem.

\vskip 2truecm

\leftline{\bf 5. Periodic Solutions of the KG Equation}

\vskip 1truecm

In this section we explicitly want to show the consistency  of BC
with the general solution of the KG equation.  We concentrate  on
the case of periodic BC.

\vskip 1truecm

\leftline{\sl A.  Explicit  Solution via Fourier-Laplace
Transformation}

\vskip 1truecm

Our starting point is the Hamiltonian version of the KG equation

$$  \dot   \phi   (x^+,   x^-)   =  -\frac{m^2}{4}   \intl   dy^-
\left[\sfrac{1}{2} \sgn (x^- - y^-) - \frac{x^- - y^-}{2L}\right]
\phi (x^+, y^-) \; , \tag 5.1  $$

where  $\phi$  is periodic  in $x^-$ and $\phi(x_0^+  = 0, x^-) =
g(x^-)$.  Equation (5.1) is an integral differential equation for
$\phi$. To solve it we introduce the Laplace transform

$$\Phi (p, x^-) = \int_0^\infty dx^+ \e^{-px^+}\phi (x^+, x^-) \;
; \quad Re(p) > 0 \; . \tag 5.2 $$

Laplace transforming (5.1) yields

$$ \Phi (p, x^-) = \frac{1}{p} g(x^-) - \frac{m^2}{4p} \intl dy^-
\left[\sfrac{1}{2}  \sgn(x^- - y^-) - \frac{x^- -y^-}{2L} \right]
\Phi (p, y^-) \; . \tag 5.3  $$

This  is  an  inhomogeneous   Fredholm  equation  of  the  second
kind\ref{33}  for $\Phi(p, x^-)$ in the variable $x^-$, which can
be solved by Fourier  transformation  with respect  to $x^-$.  To
this end we expand

$$ \eqalignno{\phi  (x^+, x^-) &= \sum_n  \phi_n (x^+) \e^{-in\pi
x^-/L} \; , & (5.4a) \cr
g(x^-) &= \sum_n g_n \e^{-in\pi x^-/L} \; , & (5.4b) \cr
\Phi(p, x^-) &= \int_0^\infty  dx^+ \e^{-px^+} \sum_n \phi_n(x^+)
\e^{-in\pi  x^-/L} =:  \sum_n \Phi_n(p) \e^{-in\pi  x^-/L} \; , &
(5.4c) \cr} $$

where we have introduced the discretised momenta\ref{7}

$$  k_n^+ = 2\pi n /L \; . \tag 5.5  $$

The Green function has the Fourier series representation

$$  \sfrac{1}{2}   \sgn(x^-  -  y^-)  -  \frac{x^-  -y^-}{2L}   =
\frac{i}{L}  \sum_{n  \ne  0}  \frac{1}{k_n^+}\e^{-in\pi  (x^-  -
y^-)/L} \; , \tag 5.6  $$

and it is obvious that its zero mode vanishes. Inserting (5.4$a$)
into the KG equation immediately yields

$$ \phi_{n=0} (x^+) = 0 \; , \tag 5.7  $$

which is nothing  but the statement  that the zero mode of a free
scalar field vanishes.   The sums in (5.4) therefore  run over $n
\ne 0$ only. Fourier transforming (5.3) then yields the solution

$$ \Phi_n(p) \equiv \int_0^\infty  dx^+ \e^{-px^+} \phi_n(x^+)  =
\frac{g_n}{p + im^2 L/4\pi n} \; ; \quad n \ne 0\; . \tag 5.8 $$

Performing  an  inverse  Laplace  transformation  gives  via  the
residue technique

$$  \phi_n(x^+)  = \frac{1}{2\pi  i} \int_{c-i\infty}^{c+i\infty}
\e^{px^+}\Phi_n(p)  = g_n \exp (-i \hat k_n^- x^+/2) \; , \tag 5.9
$$

where we have introduced the on-shell momentum

$$ \hat k_n^- = m^2/k_n^+  = \frac{m^2 L}{2\pi n} \; .  \tag 5.10
$$

Finally, the periodic solution of the KG equation is

$$ \phi (x^+,  x^-)  = \sum_{n  \ne 0} g_n \e^{-i k_n^+ x^- /2 - i
\hat k_n^- x^+/2} \equiv \sum_{n  \ne 0} g_n \e^{-i \hat k_n \cdot
x} \; . \tag 5.11  $$

Again,  as discussed  at the  end of Section  3, equation  (5.11)
suggests  that $\phi$ is uniquely determined  by data on a single
light  front  $x^+  = 0$, given  in terms  of $g(x^-)$  {\sl and}
periodic BC holding for all $x^+$. In the next subsection we will
prove that this is indeed the case.

\vskip 1truecm

\leftline{\sl B. Relation between the Characteristic Data}

\vskip 1truecm

If the last statement is true, the data $g(x^-)$ and $f(x^+)$ can
no longer be independent.   $f$ should rather  be expressible  in
terms of $g$.  To verify this we consider the solution (3.10)  of
the characteristic  problem  and set $x_0^+  = 0$, $x_0^-  = -L$.
Demanding periodic BC results in a Fredholm equation of the first
kind\ref{33} for $f$,

$$ \int _0^{x^+} dy^+ k_L (x^+ - y^+) f(y^+) - h_L (x^+) = 0 \; ,
\tag 5.12  $$

where we have introduced the abbreviations

$$  \eqalignno{k_L  (x^+  -  y^+)  &=  \frac{1}{2(x^+  - y^+)}  m
\sqrt{2L (x^+ - y^+)} J_1 \Bigl(m \sqrt{2L (x^+ - y^+)}\Bigr) \; ,
&  (5.13a)   \cr
h_L (x^+) &= \sfrac{1}{2}  \intl  dy^- J_0 \Bigl(m\sqrt{x^+  (L -
y^+)}\Bigr) \partial_y^+ g(y^-) \; . & (5.13b) \cr} $$

Again, this can be solved via Laplace transformation. Introducing
the transforms

$$\eqalignno{K_L(p)  &=  \int  dx^+  \e^{-px^+}  k_L(x^+)  \;  , &
(5.14a) \cr
F(p)  &=  \int  dx^+  \e^{-px^+}  f(x^+)  \;  , &
(5.14b) \cr
H_L(p)  &=  \int  dx^+  \e^{-px^+}  h_L(x^+)  \;  , &
(5.14c) \cr}  $$

one obtains the simple expression $F = H_L/K_L$ or

$$ f(x^+)  = \frac{1}{2\pi  i} \int_{c-i\infty}^{c+i  \infty}  dp
\, \e^{px^+} \frac{H_L(p)}{K_L(p)} \; ; \quad x^+ > 0 \; . \tag 5.15
$$

$K_L$ is given by the integral

$$ K_L(p) = \int_0^\infty  dz \, \e^{-pz^2/2Lm^2}  J_1 (z) \; , \tag
5.16  $$

which can be evaluated\ref{34} to give

$$ K_L(p) = 1 - \exp (-m^2 L/2p) \; , \tag 5.17  $$

so that finally $f$ is determined as

$$ f(x^+) = \frac{1}{2\pi  i} \int_{c-i\infty}^{c+i  \infty}  dp
\, \e^{px^+} \frac{H_L(p)}{1  - \exp (-m^2 L/2p)} \; ; \quad x^+ > 0
\; . \tag 5.18  $$

As $H_L$  is a known  functional  of $g$, we have shown  that the
data $f$ on $x^- = -L$ are given in terms of the data on $x^+ =0$
and  cannot  be specified  independently.\abs
We note that the method  breaks  down  in the limit  of vanishing
mass, $m \to 0$. In this case $k_L$ is zero and (5.12) reduces to
$h_L (x^+) = 0$, which from (5.13$b$)  is nothing but the trivial
statement that $g$ is periodic in $x^-$.  This breakdown can also
be seen by directly  studying  the Riemann solution (3.10) in the
massless case.  Periodic BC do not lead to a relation between the
data.  Antiperiodic  BC,  however,  are  more  restrictive:  they
eliminate  right  moving  waves, $f(x^+)  = 0$, so that  $\phi  =
g(x^-)$ only.

\vskip 1truecm

\leftline{\sl C. The Periodic Pauli-Jordan Function}

\vskip 1truecm

In this subsection  we want to check  our formula  (5.18)  for an
explicit  solution  of the KG equation  with known values on both
characteristics, namely the (periodic) Pauli-Jordan function.  We
start from the Fourier  representation  (2.8) of the Pauli-Jordan
function and replace

$$ \int dk^+ \to \frac{2\pi}{L}  \sum_{n  \in \bbs{Z}} \; .  \tag
5.19 $$

This yields the periodic Pauli-Jordan  function  $\Delta_L$  in a
finite spatial volume,

$$ \eqalign{\Delta_L  (x^+,  x^-) &= - \frac{i}{2L}  \sum_n  \int
dp^- \delta(p_n^+ p^- - m^2) \sgn(p_n^+ + p^-) \e^{-ip_n^+ x^- \! /2
- ip^- x^+ \! /2} \cr
&\equiv \sum_n \Delta_n (x^+) \e^{-in\pi x^- / L} \; .  \cr} \tag
5.20  $$

As $\Delta_L$  is a periodic solution of the KG equation its zero
mode must vanish, according to (5.7), i.e.

$$ \Delta_0 (x^+) = 0 \; . \tag 5.21  $$

The sum in (5.20) therefore only runs over $n \ne 0$. Thus we can
unambiguously perform the $p^-$ integration (cf.~(3.17)),

$$ \eqalign{\Delta_{n  \ne 0} (x^+)  &= -\frac{i}{2L}  \int  dp^-
\frac{1}{\vert  p_n^+  \vert}  \delta  (p^-  - m^2/p_n^+)  \,\sgn
(p_n^+ + p^-) \, \e^{-i p^- x^+ /2} = \cr
&= -\frac{i}{4\pi n} \exp (-i \hat p_n^- x^+ / 2) \; .  \cr} \tag
5.22 $$

Note that the expression  $1/\vert  p_n^+ \vert$  is always  well
defined because $p_n^+ \ne 0$. Finally we obtain

$$ \Delta_L (x^+ , x^-) = -\sum_{n \ne 0} \frac{i}{4\pi  n} \e^{-i
\hat p_n^- x^+ /2 - in\pi x^- /L} \; . \tag 5.23  $$

This coincides with formula (5.11) applied to the special case of
the Pauli-Jordan function. On the characteristics, $\Delta_L$ has
the values

$$  \eqalignno{\Delta_L  (x^+  = 0,  x^-)  &=  - \sum_{n  \ne  0}
\frac{i}{4\pi n} \e^{-i n \pi x^- /L} = - \Bigl[\sfrac{1}{4} \sgn
(x^-) - \frac{x^-}{4L}\Bigr] \equiv g (x^-) \; , & (5.24a) \cr
\Delta_L  (x^+,  x^-  = -L) &= - \frac{i}{4\pi}  \sum_{n  \ne  0}
\frac{(-1)^n}{n}  \exp \Bigl(-i  \frac{m^2  L}{4\pi n} x^+ \Bigr)
\equiv f(x^+) \; . & (5.24b) \cr} $$

As a cross-check we note that $\Delta_L (x^+ = 0, x^-)$ coincides
(up to a factor of $-2$) with the periodic  Green function  $G_3$
of (4.24). The commutators corresponding to (5.24) are

$$ \eqalignno{&[\phi  (x), \phi (0)]_{x^+  =0} = i g(x^-)  \; , &
(5.25a) \cr
&[\phi  (x), \phi (0)]_{x^-  =-L} = i f(x^+)  \; . &
(5.25b) \cr} $$

Our objective  is to obtain  (5.24$b$)  from  (5.24$a$)  by using
formula  (5.18)  so that the commutator  (5.25$b$)  on $x^- = -L$
can be expressed  in terms of the commutator  (5.25$a$) on $x^+ =
0$. To this end we calculate (using the notation of Subsection B)

$$ h_L (x^+)  = -\sfrac{1}{2}  J_0  \Bigl(m  \sqrt{x^+L}\Bigr)  +
\frac{J_1 \Bigl(m \sqrt{2Lx^+}\Bigr)}{m  \sqrt{2L x^+}} \; , \tag
5.26 $$

which has the Laplace transform

$$ H_L(p) = -\frac{1}{2p}  \exp \Bigl(-\frac{m^2L}{4p}  \Bigr)  -
\frac{1}{m^2   L}  \exp  \Bigl(  -\frac{m^2   L}{2p}   \Bigr)   +
\frac{1}{m^2 L} \; . \tag 5.27 $$

Inserting this into (5.18) one finds that $f$ should be

$$ f(x^+)  {\buildrel  ?  \over = } -\frac{1}{8  \pi i} \int_{c  -
i\infty}^{c  + i\infty}  dp \, \e^{px^+}  \left[\frac{1}{p  \sinh
(m^2L / 4p)} - \frac{1}{m^2L}\right] \; . \tag 5.28 $$

For $x^+ > 0$ the second term does not contribute. The first term
has single poles at

$$ p_n := -i\frac{m^2  L}{4n  \pi}  \; ; \quad  n= \pm 1, \pm  2,
\ldots \; . \tag 5.29  $$

Note that there is no pole at $n = 0$.  A residue  evaluation  of
(5.28) yields

$$  - \frac{1}{8  \pi  i} \!\!   \int\limits_{c  - i\infty}^{c  +
i\infty} \!\!\!  dp \, \e^{px^+}  \left[\frac{1}{p  \sinh (m^2L /
4p)} - \frac{1}{m^2L}\right]  = - \frac{i}{4\pi}  \sum_{n  \ne 0}
\frac{(-1)^n}{n}  \exp \Bigl(-i  \frac{m^2  L}{4\pi n} x^+ \Bigr)
\equiv f(x^+) \; . \tag 5.30 $$

This coincides exactly with (5.24$b$)  which is what we wanted to
show.  As a final step we want to derive a consistency  condition
analogous to (2.12) and (3.15), expressing the commutator  $[\phi
(x), \phi (0)]$  in terms of its values  on $x^+ = 0$ only.  This
information is implicitly provided by (5.23) from which we obtain

$$ [\phi  (x),  \phi  (0)]  = i \Delta_L  (x) = \sum_{n  \ne 0} i
\Delta_n (x^+ = 0) \e^{- i \hat p_n \cdot x} \; , \tag 5.31  $$

with $\Delta_n$ given in terms of the commutator on $x^+ = 0$,

$$ i\Delta_n  (x^+ = 0) = \frac{1}{2L}\intl  dx^- \e^{in\pi x^- /L}
[\phi (x), \phi (0)]_{x^+ = 0} \; . \tag 5.32  $$

Inserting this into (5.31) finally yields

$$ \eqalign{  [\phi  (x) , \phi (0)] &= \frac{1}{2L}  \intl  dy^-
[\phi (y), \phi(0)]_{y^+  = 0} \sum_{n  \ne 0} \e^{-in\pi  (x^- -
y^-)/L - i\hat p_n^- x^+ /2} = \cr
&= \intl dy^- [\phi (y) , \phi (0)] \left.   \pd{}{y^-}  \Delta_L
(x-y) \right\vert_{y^+ = 0} \; . \cr} \tag 5.33  $$

This   is   the   desired   consistency    condition    for   the
initial-boundary  value problem.  An equal-$x^-$ commutator as in
(3.15) does not appear.

\vskip 2 truecm

\leftline{\bf 6. Conclusions}

\vskip 1truecm

We have shown that a single-time  formalism with the canonical LC
Hamiltonian and data/commutators specified on {\sl two} lightlike
hyperplanes  $x^\pm  = const$ fails to describe  the dynamics  in
accordance with the Euler-Lagrange equation of motions.  However,
quantisation on a {\sl single} light front (characteristic)  $x^+
= const$  is consistent  if (i) the field under consideration  is
massive and (ii) additional boundary conditions  are specified in
the $x^-$ direction.  Hamiltonian and Euler-Lagrange  description
of  the  dynamics  are  then  completely   equivalent.    The  LC
Hamiltonian formulation is canonical and simple.  Technically, we
have found that the phase space reduction  method of Faddeev  and
Jackiw is a very useful tool for these investigations  as well as
for light front quantisation  in general.  As a result we can say
that  (under  the above  mentioned  conditions)  the attempts  of
quantising  LC field  theories  in finite  volumes  are on a safe
basis.\abs
One might argue that our conclusions  are valid only for the free
theory.  However,  as has been emphasized  by Jackiw\ref{4},  the
equal-$x^+$-commutators are also initial data for the interacting
theory,  as long  as there  is no dynamics  evolving  within  the
initial surface. Due to causality, this is obviously the case for
spacelike  surfaces  but also for lightlike  ones, if there  is a
mass gap. Again the massless case seems to be peculiar.\abs There
is no such staightforward  argument concerning  the relevance  of
boundary conditions  for interacting  theories,  but there are at
least some indications.   For theories  with constant  background
fields, caused e.g.~by non-vanishing  vacuum expectation  values,
periodic boundary conditions seem to be a natural choice\ref{13}.
Quite generally one would guess that boundary conditions are less
important  in  the  perturbative  than  in  the  non-perturbative
regime.   This  has been  verified  for LC-$\phi^4$-theory  where
perturbative contributions  to the self energy caused by boundary
effects   were   shown   to  vanish   in  the   infinite   volume
limit\ref{35}.  On the other hand, if non-perturbative physics is
due to non-trivial  topological  properties,  boundary conditions
play a crucial  role.   This is well known  from the attempts  of
quantising  gauge  theories  in  finite  volumes  where  boundary
conditions define additional topological quantum numbers\ref{36}.
In the context of LC quantum field theory this has been discussed
for the Schwinger  model\ref{37,38}.   For LC gauge theories with
massless gauge bosons (no Higgs phenomenon),  however, one has to
clarify the issue of massless  particles  propagating  within the
quantisation surface.\abs
This implies that one should extend the analysis of this paper to
fields  of higher spin (fermions,  gauge bosons,...)  and also to
higher dimensions.  To some extent this has been done in Ref.[15]
(without discussion  of boundary conditions)  where it is claimed
that the characteristic  initial  value problem  "is undetermined
for field theories containing  massive fields of spin higher than
1/2".  If  true  there  might  also  be  a  problem  for  the  LC
formulation  of massive  gauge bosons acquiring  their masses via
the Higgs mechanism.   This raises  the general  issue of a light
front description  of mass generation:   The conclusions  of this
paper  only  hold  for  massive  fields;  the  massless  case  is
generically  different.   There is probably  no smooth transition
between  the massless  and the massive  case with respect  to the
quantisation  prescription.   This  can  be seen  already  in the
Schwinger  model  (QED  in  1+1  dimensions).    There,  massless
fermions,  which have to be quantised  on the light cone\ref{16},
transform  into  a massive  boson,  which  can be quantised  on a
single front with periodic  boundary  conditions\ref{37}.   It is
therefore very difficult to set up a direct bosonisation  formula
because  the latter has to relate fields  quantised  on different
hypersurfaces.\abs
In higher  dimensions  it is clear  that there  is a much greater
variety    of   initial    and/or   boundary    conditions    and
solutions\ref{25, 39}.  For massive fields a formula analogous to
(3.8)  still  holds  with just additional  integrations  over the
additional dimensions\ref{14}.  The discussion of sections 3 to 5
should therefore (with minor modifications) still be valid.

\vfill\eject

\leftline{\bf Acknowledgements}

\vskip 1truecm

We are indebted  to M.J.~Lavelle,  G.~Fischer  and S.~Simb\"urger
for reading (versions  of) the manuscript.

\vskip 1truecm

\leftline{\bf References}

\vskip 1truecm

{\parindent=1truecm

\litem{[1]}
P.A.M. Dirac, Rev. Mod. Phys. {\bf 21} (1949) 392
\litem{[2]}
S. Fubini, G. Furlan, Physics {\bf 1} (1965) 229
\litem{}
S. Weinberg, Phys. Rev. {\bf 150} (1966) 1313
\litem{}
J. Jersak, J. Stern, Nucl. Phys. {\bf B7} (1968) 413
\litem{}
H.  Leutwyler, in: Springer Tracts in Modern Physics, Vol.~50, G.
H\"ohler, ed., Springer, Berlin, Heidelberg, New York 1969
\litem{[3]}
J.D. Bjorken, Phys. Rev. {\bf 179} (1969) 1547
\litem{}
S.D.  Drell, D.  Levy, T.M.  Yan, Phys.  Rev.  {\bf 187} (1969)
2159; {\bf D1} (1970) 1035
\litem{[4]}
R.  Jackiw, in:  Springer Tracts in Modern Physics, G.  H\"ohler,
ed., Springer, Berlin, Heidelberg, New York 1972
\litem{[5]}
S.J. Chang, S.K. Ma, Phys. Rev. {\bf 180} (1969) 1506
\litem{}
J.B. Kogut, D.E. Soper, Phys. Rev. {\bf D1} (1970) 2901
\litem{}
J.D.  Bjorken, J.B.  Kogut, D.E.  Soper, Phys.  Rev.  {\bf D3}
(1971) 1382
\litem{}
R.A. Neville, F. Rohrlich, Phys. Rev. {\bf D3} (1971) 1692
\litem{[6]}
S.J.  Chang, R.G.  Root, T.M.  Yan, Phys. Rev. {\bf D7} (1973)
1133
\litem{}
S.J.  Chang, T.M.  Yan, Phys. Rev. {\bf D7} (1973) 1147
\litem{}
T.M.  Yan, Phys. Rev. {\bf D7} (1973) 1760, 1780
\litem{[7]}
H.C.  Pauli, S.J.  Brodsky, Phys.  Rev.  {\bf D32} (1985) 1993,
2001
\litem{}
T. Eller, H.C. Pauli, S.J. Brodsky, Phys. Rev. {\bf D35} (1987)
1439
\litem{}
K.  Hornbostel, H.C.  Pauli, S.J. Brodsky, Phys. Rev. {\bf D41}
(1990) 3814
\litem{}
A.C.  Tang, H.C.  Pauli, S.J.  Brodsky, Phys.  Rev.  {\bf D44}
(1991) 1842
\litem{[8]}
R.J. Perry, A. Harindranath, K.G. Wilson, Phys. Rev. Lett. {\bf
65} (1990) 2959
\litem{}
R.J.  Perry, A.  Harindranath, Phys.  Rev. {\bf D43} (1991) 492,
4051
\litem{}
D.  Mustaki, S. Pinsky, J. Shigemitsu, K. Wilson, Phys. Rev. {\bf
D43} (1991) 3411
\litem{}
D. Mustaki, S. Pinsky, Phys. Rev. {\bf D45} (1992) 3775
\litem{[9]}
H.  Leutwyler,  J.R.  Klauder, L.  Streit, Nuovo Cim.  {\bf 66A}
(1970) 536
\litem{[10]}
G.P. Lepage, S.J. Brodsky, T. Huang, P.B. Mackenzie, Banff Summer
Institute   on  Particle   Physics,   Banff,   Alberta,   Canada,
CLNS-82/522, 1981
\litem{[11]}
J.M.  Namys\l owski, Prog.  Part. Nucl. Phys. {\bf 14} (1984) 49,
\litem{[12]}
C.  Dietmaier, T.  Heinzl, M.  Schaden, E.  Werner, Z. Phys. {\bf
A334} (1989) 215
\litem{}
P.P.  Srivastava,  {\sl Spontaneous  Symmetry  Breaking in Light
Front Quantized Field Theory}, OSU preprint (1991)
\litem{}
S. Huang, W. Lin, {\sl On the Zero Mode Problem of the Light-Cone
Quantization}, University of Washington preprint (1992)
\litem{}
D.   Robertson,   {\sl  On  Spontaneous   Symmetry  Breaking   in
Discretized Light-Cone Field Theory}, preprint SMUHEP/92-3 (1992)
\litem{}
C.M.  Bender, S.  Pinsky, B.  van de Sande, Phys.  Rev. {\bf D48}
(1993) 816
\litem{}
M. Maeno, {\sl $k^+ = 0$ Modes in Light Cone Quantization}, Osaka
preprint, OU-HET-175B (1993)
\litem{[13]} T.  Heinzl, S.  Krusche, E.  Werner, Phys. Lett. {\bf
B272} (1991) 54
\litem{}
T.  Heinzl, S.  Krusche, S. Simb\"urger, E. Werner, Z. Phys. {\bf
C56} (1992) 415
\litem{[14]}
R.A. Neville, F. Rohrlich, Nuovo Cim. {\bf A1} (1971) 625
\litem{}
F. Rohrlich, Acta Physica Austriaca, Suppl. VIII (1971) 277
\litem{[15]}
G.  Domokos, in:  Boulder Lectures, Vol.  XIV, 1971, A.O. Barut,
W.E. Brittin eds., Colorado University Press, Boulder 1972
\litem{[16]}
G. McCartor, Z. Phys. {\bf C41} (1988) 271
\litem{[17]}
P.A.M. Dirac, Canad. J. Phys. {\bf 1} (1950) 1
\litem{}
P.A.M.  Dirac, Lectures on Quantum Mechanics,  Benjamin, New York
1964
\litem{}
J.L. Anderson, P.G. Bergmann, Phys. Rev. {\bf 83} (1951) 1018
\litem{}
P.G. Bergmann, Helv. Phys. Acta Suppl. IV (1956) 79
\litem{[18]}
K.  Sundermeyer,  Constrained  Dynamics, Lecture Notes in Physics
{\bf 169}, Springer, Berlin, Heidelberg, New York 1982
\litem{[19]}
L. Faddeev, R. Jackiw, Phys. Rev. Lett. {\bf 60} (1988) 1692
\litem{}
R.  Jackiw, {\sl (Constrained)  Quantisation  Without Tears}, MIT
preprint CTP \# 2215, hep-th/9306075 (1993)
\litem{[20]}
P. Jordan, W. Pauli, Z. Phys. {\bf 47} (1928) 151
\litem{[21]}
J.  Schwinger,  Phys. Rev.  {\bf 75} (1949)  651, appendix
\litem{[22]}
R. Courant, D. Hilbert, Methods of Mathematical Physics, Vol.~II,
Interscience, New York 1962
\litem{}
P.R.  Garabedian, Partial Differential Equations, Wiley and Sons,
New York, London, Sidney, 1964
\litem{[23]}
T.  Myint-U, L. Debnath,
Partial  Differential  Equations  for  Scientists  and Engineers,
North Holland, New York, Amsterdam, London 1987
\litem{[24]}
S.  Schweber,  An  Introduction  to  Relativistic  Quantum  Field
Theory, Chapter 7c, Harper and Row, New York 1961,
\litem{[25]}
N.E.   Ligterink,   B.L.G.   Bakker,  {\sl  Partial  Differential
Equations  and  Null  Plane  Field  Theory},  Vrije  Universiteit
Amsterdam preprint (1993)
\litem{[26]}
N.N.  Bogolubov, A.A.  Logunov, A.I. Oksak, I.T. Todorov, General
Principles  of Quantum Field Theory, Chapter 11, Kluwer Academic,
Dordrecht, Boston, London 1990
\litem{[27]}
O.C.  Jacob,  {\sl Parity  and Front-Form  Quantisation  of Field
Theories}, Stanford preprint 6391 (1993)
\litem{}
O.C.  Jacob,  {\sl Quantization  of Gauge Field  Theories  on the
Front-Form  without  Gauge  Constraints  I:  The  Abelian  Case},
Stanford preprint 6392 (1993)
\litem{[28]}
I.M.   Gelfand,  G.E.   Shilov,  Generalized  Functions,  Vol.~I,
Academic Press, New York 1964
\litem{[29]}
S. Schlieder, E. Seiler, Comm. Math. Phys. {\bf 25} (1972) 62
\litem{[30]}
W.-M.  Zhang, A.  Harindranath, {\sl Light-Front QCD:  I. Role of
Longitudinal   Boundary  Integrals},   OSU  preprint,   OSU-NT-\#
93-0122 (1993)
\litem{[31]}
L.  Banyai, L.  Mezinescu, Rev. Roum. Phys. {\bf 18} (1973) 1035;
Phys. Rev. {\bf D8} (1973) 417
\litem{}
P. Senjanovi\v c, Ann. Phys. (N.Y.) {\bf 100} (1976) 227
\litem{[32]}
T.  Suzuki, S.  Tameike, E.  Yamada, Progr. Theor. Phys. {\bf 55}
(1976) 922
\litem{}
T.  Maskawa, T. Yamawaki, Progr. Theor. Phys. {\bf 56} (1976) 270
\litem{}
M.  Huszar, J.  Phys.  A: Math. Gen. {\bf 9} (1976) 1359
\litem{}
P.J.  Steinhardt,  Ann.  Phys. (N.Y.) {\bf  128}  (1980)  425
\litem{[33]}
P.M.  Morse, H. Feshbach, Methods of Theoretical Physics, Part I,
McGraw-Hill, New York 1953
\litem{[34]}
I.S.  Gradshteyn,  I.M.  Ryzhik,  Table of Integrals,  Series and
Products, Formula 6.618.1, Academic, New York 1980,
\litem{[35]}
T.  Heinzl, S.  Krusche, E. Werner, B. Zellermann, {\sl Zero Mode
Corrections  in Perturbative  Light  Cone Scalar  Field  Theory},
preprint TPR 92-17 (1992)
\litem{[36]}
G. 't Hooft, Nucl. Phys. {\bf B153} (1979) 141
\litem{}
M. L\"uscher, Nucl. Phys. {\bf B219} (1983) 233
\litem{}
P. van Baal, J. Koller, Ann. Phys. (N.Y.) {\bf 174} (1987) 299
\litem{}
P.   van  Baal,  {\sl  A Review  of QCD  Spectroscopy  in  Finite
Volumes},  talk presented  at the Third Light Cone Meeting,  PSI,
Villigen, Switzerland, 1993
\litem{[37]}
T.  Heinzl, S.  Krusche, E. Werner, Phys. Lett. {\bf B256} (1991)
55
\litem{[38]}
T.  Heinzl, S.  Krusche, E. Werner, Phys. Lett. {\bf B275} (1992)
410
\litem{[39]}
F.G. Friedlander, Proc. Roy. Soc. {\bf A269} (1962) 53
\litem{} P.
Hillion, J. Math. Phys. {\bf 31} (1990) 1939
}

\vfill\eject

\leftline{\bf Figure Captions}

\vskip 1truecm

{\parindent=2truecm

\litem{Figure 1:}
The integration contour used in the derivation of (3.8).  To know
the  field  $\phi$  at the point  $P$ one has  to specify  either
$\phi$ and $\dot \phi$ on $AB$ (Cauchy problem) or $\phi$ on $AC$
{\sl and} $CB$ (CIVP).  Data outside  the characteristic  cone of
$P$ need not be specified\ref{14, 23}.

\vskip 1truecm

\bye